\documentclass[11pt]{article}
\pdfoutput=1
\usepackage{jheppub}
\usepackage{tabu}
\usepackage[vcentermath]{youngtab}
\usepackage[usenames,dvipsnames,table]{xcolor}
\usepackage{graphicx,amsmath,amssymb,amsthm,multirow,array,bm,bbm,esint}
\usepackage[mathscr]{eucal}
\usepackage[bbgreekl]{mathbbol}  
\usepackage{epsf,amsfonts}
\usepackage{slashed}
\usepackage{ulem}
\usepackage[numbers,sort&compress]{natbib}
\usepackage{minibox}
\usepackage{rotating}
\usepackage{pdflscape}
\usepackage{array,tikz-cd}
\usetikzlibrary{decorations.pathmorphing}
\usetikzlibrary{decorations.pathreplacing}

\tikzset{snake it/.style={decorate, decoration=snake}}

\usepackage{subfig}
\usepackage{float}
\definecolor{rust}{rgb}{0.8,0.2,0.2}

\usepackage{soul,xcolor}
\setstcolor{red}

\def\be{\begin{equation}}
\def\ee{\end{equation}}
\def\bea{\begin{eqnarray}}
\def\eea{\end{eqnarray}}
\def\ie{\begin{equation}\begin{aligned}}
\def\fe{\end{aligned}\end{equation}}
\def\ba{\begin{aligned}}
\def\ea{\end{aligned}}
\newcommand{\vev}[1]{{\left< {#1} \right>}}
\newcommand{\bra}[1]{{\left< {#1} \right|}}
\newcommand{\ket}[1]{{\left| {#1} \right>}}

\def\bC {\mathbb{C}}
\def\bZ {\mathbb{Z}}

\newcommand{\m}{\mu}
\newcommand{\n}{\nu}
\newcommand{\A}{{\alpha}}
\newcommand{\B}{{\beta}}

\newcommand{\ca}{{\mathfrak{c}}}

\DeclareMathOperator{\erf}{erf}
\DeclareMathOperator{\erfc}{erfc}
\DeclareMathOperator{\Res}{Res}

\title{Spinning constraints on chaotic large $c$ CFTs}

\author{Chi-Ming~Chang, David M.~Ramirez, Mukund~Rangamani}
\affiliation[]{
Center for Quantum Mathematics and Physics (QMAP),  \\
Department of Physics, University of California, Davis, CA 95616 USA.}

\emailAdd{wychang@ucdavis.edu, dramir@ucdavis.edu, mukund@physics.ucdavis.edu}

\abstract{We study out-of-time ordered four-point functions in two dimensional conformal field theories by suitably analytically continuing the Euclidean correlator.  For large central charge theories with a sparse spectrum, chaotic dynamics is revealed in an exponential decay; this is seen directly in the contribution of the vacuum block to the correlation function. However, contributions from individual non-vacuum blocks with large spin and small twist dominate over the vacuum block. We argue, based on holographic intuition, that suitable summations over such intermediate states in the block decomposition of the correlator should  be sub-dominant, and attempt to use this criterion to constrain the OPE data with partial success. Along the way we also discuss the relation between the spinning Virasoro blocks and the on-shell worldline action of spinning particles in an asymptotically AdS spacetime.
}

\begin{document}
\maketitle


\section{Introduction}
\label{sec:intro}

It is empirically clear that field theories with a large number of degrees of freedom (measured e.g., by the central charge) and a sparse spectrum of low-lying operators  satisfy necessary criteria to have a dual description in terms of gravitational dynamics in AdS. While it has been conjectured that these criteria are also sufficient \cite{Heemskerk:2009pn}, it is far from obvious that these criteria alone would suffice. For one, the spectral information is a rather crude characterization of the field theory; one is still missing information about the operator algebra, encoded for example in the operator product expansion, which tells us what the relative likelihood of physical interactions in the theory is.

To proceed further, one needs to be a bit more specific about what is meant for a theory to be holographic. The large central charge $c \gg 1$  is supposed to serve in general as a proxy for a suitable planar expansion, with an effective Planck's constant $\hbar/c$, guaranteeing a classical description (in a saddle point sense). This classical description, per se, does not have to be local; it could be a classical string theory with a finite string scale $\ell_s$. For instance, the symmetric product of $N$ copies of the compact free boson in $(1+1)$-dimensions  has a central charge $c\sim N$ which can be made large, and satisfies the sparseness criterion \cite{Hartman:2014oaa}, but is expected to be dual to a classical (tensionless) string  theory at the orbifold point (cf., \cite{Gaberdiel:2018rqv,Giribet:2018ada}). With sufficient supersymmetry there is a moduli space of vacua, wandering off along which is expected to bring one to a point where the string theory description collapses onto classical supergravity, but this is non-generic.

One can therefore ask, what are the quantifiable features that allow a given field theory to admit a classical gravitational holographic description, preferably with a two derivative gravitational action, a la Einstein-Hilbert. While we do not yet have a ready answer to this question, many groups have tried to sharpen this by examining various observables, with the hope of extracting the essential features. In what follows, we will examine a particular observable, the chaos correlator in $(1+1)$-dimensional CFTs (henceforth CFT$_2$),  to gain some insight into this question.

The chaos correlator is an out-of-time-order (OTO) correlation function probing a thermal state (at temperature $\beta^{-1}$), of the form 
$\mathcal{C}(t,x) = \langle W(t,x) \, V(0,0) \, W(t,x) \, V(0,0) \rangle_\beta$ aimed at measuring how a change in initial conditions propagates in  quantum state. For ergodic field theories, it is argued that  $\mathcal{C}(t,x)$ decays exponentially around the scrambling time $t_* \sim \beta \, \log c $, viz., 
\ie\label{eqn:exp_decay}
\mathcal{C}(t,x) \sim  \mathcal{C}_0 - \mathcal{C}_1  \, e^{\lambda_L \, (t-t_*) } + \cdots.
\fe
This behaviour is seen in holographic examples where black hole physics dictates that the Lyapunov exponent $\lambda_L = \frac{2\pi}{\beta}$. In a seminal paper \cite{Maldacena:2015waa}, it was argued that the holographic answer arising from black holes is an upper bound and in general $\lambda_L \leq \frac{2\pi}{\beta}$ with black holes  attaining the optimal value.\footnote{ Another situation where the bound is saturated is the SYK model \cite{Kitaev:2015aa,Maldacena:2016hyu,Kitaev:2017awl} which involved random all-to-all couplings of Majorana fermions in a quantum mechanical setting.} The arguments for the bound rely on analytic properties of thermal correlation functions. Inclusion of string theory effects is expected to lower the Lyapunov exponent, with 
$\lambda_L = \frac{2\pi}{\beta}  \left(1- \gamma_s \, \frac{\ell_s^2}{R^2}\right) $ for some $\gamma_s \sim \mathcal{O}(1)$ and $R$ being a characteristic curvature length scale \cite{Shenker:2014cwa}. 

Let us now take stock of what is known about the chaos correlator in field theories. As mentioned earlier we have quantum mechanical models like SYK which saturate the bound. In $(1+1)$ dimensions, there are some  generalized  SYK models which do not saturate the bound, but rather attain $\lambda_L \sim 0.6\, \frac{2\pi}{\beta} $ \cite{Murugan:2017eto,Bulycheva:2018qcp,Peng:2018zap} which is consistent with  a classical  string holographic dual. The symmetric product orbifold alluded to earlier however gives $\lambda_L =0$ belying its integrable nature at the orbifold point \cite{Perlmutter:2016pkf}.

Of interest to us however is the set-up considered in \cite{Roberts:2014ifa}  who focused on CFT$_2$s with large central charge. 
Specifically, they examined the Euclidean correlation function $\langle V V WW \rangle_\beta$  in an  OPE channel $V\, V \to \mathcal{O}_h\to WW$, with intermediate operator $\mathcal{O}_h $. By truncating the correlator to include only the identity operator $\mathcal{O}_0 = \mathbb{1}$ and its Virasoro descendants, which assumes the low-lying spectrum  is indeed sparse, they were able to show that upon analytic continuation to the appropriate Lorentzian out-of-time-order, one obtains the desired form of the chaos correlator with maximal Lyapunov exponent. This is due to a delicate cancellation between the contribution from Virasoro descendants of different spins, since each spin-$s$ global primary operator  contributes to the OTO correlator an exponential factor $e^{{2\pi\over \beta}(s-1)t}$, which for $s>2$ violates the chaos bound. At first sight, this seems reasonable, for the truncation to the Virasoro vacuum block contribution is tantamount to focusing on graviton exchange in the bulk and ignoring others.  A moment's reflection however reveals a problem: the cross ratio of the OTO four-point function at the scrambling time $t_*$ is exponentially close to the boundary of the radius of convergence of the operator product expansion (OPE).\footnote{ This is most apparent in the pillow coordinate $q$ reviewed in \S\ref{sec:pillow}.} One expects the contributions from the non-vacuum blocks to the four-point function would become important. Furthermore,  it does not follow that in the physical result only the graviton exchange dominates (the latter would require that in the $VW$ OPE can be truncated to sole vacuum contribution, which will not hold universally). Subsequent investigations have focused on subleading corrections \cite{Fitzpatrick:2016thx} while more recent work \cite{Liu:2018iki,Hampapura:2018otw} explores the limitations of the vacuum block truncation ansatz for the chaos correlator.\footnote{ The exponential structure \eqref{eqn:exp_decay} of the OTO correlators can already be seen in a further truncated sector that includes only the identity operator and the stress tensor. The contribution from the stress tensor is responsible for the exponential factor in the second term of \eqref{eqn:exp_decay}. This inspires recent attempts at constructing an effective description for chaotic dynamics \cite{Blake:2017ris}, which was specifically applied to 2d CFTs in \cite{Haehl:2018izb}.}

Our interest in the current work will be to make a careful analysis of the non-vacuum contributions in the Euclidean correlator, and implications thereof in the out-of-time-order observable of interest. We start by noting that it was already anticipated by \cite{Roberts:2014ifa} that non-vacuum blocks could in-principle have a significant contribution around the scrambling time upon analytic continuation. We will first verify this explicitly by taking into account the contributions from non-vacuum primaries. This has also been confirmed independently in  recent works \cite{Liu:2018iki,Hampapura:2018otw}. A technical aid we will employ is to use Zamolodchikov's recursion relation \cite{Zamolodchikov:1985ie,Zamolodchikov1987} to obtain the non-vacuum blocks numerically to a good accuracy (using results of \cite{Fitzpatrick:2016thx}) to enable make firm predictions. Specifically, we will see that the deviation from the vacuum block result depends on the spin of the light non-vacuum primaries in the spectrum.

Knowing the individual blocks however is insufficient for our purposes. Had there been a finite number of non-vacuum primaries, we would have trouble in estimating the chaos correlator in Lorentz signature, especially if each individual block had large contribution to the correlator. What hopefully saves the day is the fact that the CFTs of interest have an infinite number of primaries. While each is locally more important, summing over all of them with the correct OPE coefficients should result in a sub-dominant contribution. This is for example analogous to the net phase shift induced in forward scattering amplitudes in flat space, where individual higher-spin states give  more and more dominant contribution, while the resummed answer is bounded by unitarity (see eg., \cite{Camanho:2014apa}). Assuming therefore that we have an infinite number of primaries contributing, ideally we can bound their density and their OPE coefficients by demanding that the  truncation of \cite{Roberts:2014ifa} indeed gives the physically acceptable answer.  While the density of primaries will have to obey the sparseness condition, we would learn of new constraints on the OPE coefficients (which should be stronger than simple factorization statements). 

We set-up the problem of putting bounds on OPE data, and will show that the contributions from very heavy primaries and large spin primaries in the intermediate states are innocuous. This is done by explicit evaluation of the conformal blocks to get a good estimate, and uses some recently derived bounds on the growth of OPE coefficients \cite{Chang:2015qfa,Chang:2016ftb}. We are also able to estimate the contribution of the light intermediate operators, and find the need for a conspiracy between them and the moderately heavy intermediate states for the vacuum block to provide the correct answer post analytic continuation.  The somewhat sticky point of our analysis is an inability to find useful bounds for intermediate primaries which are moderately heavy $h \sim  c$. Here the problem we encounter is a technical obstacle -- we have not managed to obtain a useful estimate of the conformal blocks themselves. We do try to  exploit semi-analytic bounds on the conformal block data (see  \cite{Kusuki:2017upd, Kusuki:2018wcv, Kusuki:2018nms, Kusuki:2018wpa}) to provide an estimate, but find that the results known thus far are too limiting to get a handle on the behaviour of the blocks on timescales of order the scrambling time (they do give a handle on the very late time behaviour of the correlator).

The outline of the paper is as follows: In \S\ref{sec:review} we will review the basic features of the chaos correlator, and revisit the analysis of \cite{Roberts:2014ifa} to set the stage for the discussion. \S\ref{sec:nonvac} is devoted to demonstrating the limitations both analytically and numerically of the truncation to the vacuum block contribution to the Euclidean OPE prior to analytic continuation. In \S\ref{sec:estimates} we attempt to put bounds on the OPE data, though as advertised, our attempts will only be partially successful. Following this in \S\ref{sec:adsgrav} we take the opportunity to revisit the geodesic computation of the conformal blocks and show how spinning particles in the bulk AdS$_3$ can be used to reproduce the results derived in \S\ref{sec:nonvac} and conclude with a brief discussion in \S\ref{sec:discuss}. 
The Appendices \ref{sec:BTZshockwave} and \ref{sec:geodesics} are devoted to providing details of the spinning particle analysis, while Appendix \ref{sec:semiwline} computes the Euclidean block for completeness.

\section{Review of OTO correlators in 2d CFT}
\label{sec:review}

To set the stage for our discussion we will quickly review some salient facts about four-point functions in CFTs. We will start with the Euclidean correlation function, which can be analytically continued to the OTO regime as discussed in \cite{Roberts:2014ifa}. We will revisit some of the features of the OPE expansion and review a useful change of variables introduced in \cite{Zamolodchikov1987} which makes the analytic and convergence  properties of the correlator manifest. Finally, we discuss the chaos correlator in the limit when two of the operators are heavy and two others light, and illustrate the central claim of \cite{Roberts:2014ifa} that the truncation to the vacuum block contribution suffices to capture the Lyapunov behaviour of the correlator.

\subsection{Euclidean correlators and OTO observables}
\label{sec:euclotoc}

 Consider a Euclidean four-point function of two identical primary operators $W$ and another two identical primary operators $V$ of conformal weights $(h_w,\bar h_w)$ and $(h_v,\bar h_v)$ on a Riemann sphere $\widehat\bC\equiv\bC\cup\{\infty\}$,
\ie\label{eqn:EuclideanWWVV}
\vev{W(z_1,\bar z_1)\, W(z_2,\bar z_2)\, V(z_3,\bar z_3)\, V(z_4,\bar z_4)},
\fe
where the variables $\bar z_i$ are fixed to be the complex conjugate of $z_i$, i.e. $\bar z_i=  z_i^*$. Away from the coincident points 
$z_i=z_j$ for $i\neq j$, the four-point function is an analytic function of the variables $z_i\in \widehat\bC$.  Conformal symmetry constrains the four-point function to take the form
\ie\label{eqn:EuclideanCorrelator}
\vev{W(z_1,\bar z_1)\, W(z_2,\bar z_2)\, V(z_3,\bar z_3)\, V(z_4,\bar z_4)}=
\frac{G(z,\bar z)}{ z_{12}^{2 h_w\, }\bar z_{12}^{2\bar h_w}\, z_{34}^{2h_v\, }\bar z_{34}^{2\bar h_v}},
\fe
where $z$ and $\bar z$ are conformal cross ratios
\ie
z=\frac{z_{12}z_{34}}{ z_{13}z_{24}}\,, \qquad \bar z= \frac{\bar z_{12}\bar z_{34} }{\bar z_{13}\bar z_{24}},
\fe
which are invariants of the global conformal group SL$(2,\bC)$. The four-point function admits a decomposition in terms of the Virasoro block,
\ie\label{eqn:virasoroBlockDecomp}
G(z,\bar z)=\sum_{a} C_{WW{\cal O}_a} \, C_{VV{\cal O}_a}\, z^{2h_w}\bar z^{2h_w}{
\, \cal F}(h_w,h_v,h_a;z) \, {\cal F}(\bar h_w,\bar h_v,\bar h_a;\bar z),
\fe
where the sum runs over all the Virasoro primary operators ${\cal O}_a$ that appear in both the $W\times W$ and $V\times V$ OPEs, and $h_a$, $\bar h_a$ are the conformal weights of the Virasoro primaries ${\cal O}_a$. In the small $z$ limit, the Virasoro block ${\cal F}(h_w,h_v,h_a;z)$ has the expansion 
\ie
{\cal F}(h_w,h_v,h_a;z) = z^{h_a-2h_w} \,  \left(1+\cdots\right),
\fe
where the leading contribution comes from the primary operator ${\cal O}_a$.\footnote{ When we need to refer to a generic operator appearing the OPE expansion we will also refer to it as $\mathcal{O}$ and use $(h,\bar h)$ for its conformal weights for simplicity.} The complex plane can be conformally mapped to a cylinder by $z=e^{\frac{2\pi}{\beta}(x+i\,\tau)}$ and $\bar z=e^{ \frac{2\pi}{ \beta}(x-i\,\tau)}$, where the imaginary time coordinate $\tau$  has  periodicity $\beta$.

We are interested in the out-of-time-order (OTO) four-point function
\ie\label{OtO4-pt}
\vev{W(t,0)\, V(0,x) \, W(t,0)\, V(0,x)}_\beta\quad{\rm with}\quad t\ge x\ge0,
\fe
which is related to the Euclidean four-point function \eqref{eqn:EuclideanWWVV} by an analytic continuation.  In general, an Euclidean correlator can be analytically continued to a Lorentzian correlator by relaxing the condition $\bar z_i=  z_i^*$. As a function of the independent complex variables $z_i$ and $\bar z_i$, the correlator has singularities at either $z_i=z_j$ or $\bar z_i = \bar z_j$ for $i\neq j$, where the operators are light-like separated. There are branch cuts extending from these light-cone singularities. When analytically continuing to a Lorentzian configuration with time-like separated points, the path of the analytic continuation needs to be chosen to avoid the light-cone singularities. Different path choices give Lorentzian correlators with different operator orderings.\footnote{ A nice discussion of this can be found for example in the book \cite{Haag:1992hx} (see also \cite{Hartman:2015lfa} for a discussion in CFTs and \cite{Haehl:2017qfl} for explicit connections to OTOCs).} This is equivalent to Wick rotation with an $i\epsilon$-prescription
\ie\label{eqn:imaginary_time}
i\tau_i= t_i + i\epsilon_i,
\fe
where in our convention the operators are ordered by their ``imaginary time" $\epsilon_i$. For the case of our interest, the Euclidean four-point function \eqref{eqn:EuclideanWWVV} is Wick rotated to the OTO four-point function \eqref{OtO4-pt} by 
\ie\label{eqn:LorentzianCon}
&z_1=e^{{2\pi\over \beta}(t+i\epsilon_1)},&&z_2=e^{{2\pi\over \beta}(t+i\epsilon_2)},&&z_3=e^{{2\pi\over \beta}(x+i\epsilon_3)},&&z_4=e^{{2\pi\over \beta}(x+i\epsilon_4)},
\\
&\bar z_1=e^{-{2\pi\over \beta}(t+i\epsilon_1)},&&\bar z_2=e^{-{2\pi\over \beta}(t+i\epsilon_2)},&&\bar z_3=e^{{2\pi\over \beta}(x-i\epsilon_3)},&&\bar z_4=e^{{2\pi\over \beta}(x-i\epsilon_4)},
\fe
with imaginary times $\epsilon_i$ ordered as $\epsilon_1<\epsilon_3<\epsilon_2<\epsilon_4$. In the region $-x<t<x$, all the operators are space-like separated. At $t=x$, the operators $W$'s are at the future lightcone of the operators $V$'s.  At $t>x$, the operators $W$'s and $V$'s are time-like separated, and the ordering of the operator in the four-point function is determined by the ordering of the $\epsilon_i$.

In terms of the cross ratios $z$ and $\bar z$, the light-cone singularities at $z_1=z_2$, $z_4$, $z_3$ (or $\bar z_1=\bar z_2$, $\bar z_4$, $\bar z_3$) correspond to $z=0$, $1$, $\infty$ (or $\bar z=0$, $1$, $\infty$). The function $G(z,\bar z)$ of independent complex variables $z$ and $\bar z$ has branch cuts extending from the $z=0$, $1$, and  $\infty$, which can be chosen to lie on $(-\infty,0]$ and $[1,\infty)$. We are interested in the process when the $W$ operators approach the light-cone of the $V$ operators, which occurs at $t=\pm x$ and correspondingly at $z=1$ or $\bar z=1$. The cross ratio $z$ expanded at the light-cone $t=x$ as
\ie
z=&{\sin \pi(\epsilon_2-\epsilon_1)\sin \pi(\epsilon_4-\epsilon_3)\over \sin \pi(\epsilon_3-\epsilon_1)\sin \pi(\epsilon_4-\epsilon_2)}
\\
& \quad- i\pi (t-x){\sin\pi(\epsilon_3+\epsilon_4-\epsilon_1-\epsilon_2)\sin\pi(\epsilon_2-\epsilon_1)\sin\pi(\epsilon_4-\epsilon_3)\over \sin^2\pi(\epsilon_4-\epsilon_2)} + {\cal O}(t-x)^2.
\fe
As the time $t$ increases from $-x<t<x$ to $t>x$, the cross ratio $z$ moves across the branch cut on $[1,\infty)$ from the upper half complex plane to the lower half complex plane. By a similar analysis, one can find that as the time $t$ decreases from $-x<t<x$ to $t<-x$ the cross ratio $\bar z$ moves across the branch cut on $[1,\infty)$ from the lower half complex plane to the upper half complex plane.

\subsection{OPE convergence and pillow coordinate}
\label{sec:pillow}
\label{sec:pillow}

The four-point function with the cross ratios $z$ and $\bar z$ as independent complex variables is defined on a branched cover of the space $\widehat\bC\times\widehat\bC$. The OTO four-point function \eqref{OtO4-pt} and the Euclidean four-point function \eqref{eqn:EuclideanWWVV} are related by an analytic continuation along a path that crosses the branch cut at $z\in[1,\infty)$. The Virasoro block expansion \eqref{eqn:virasoroBlockDecomp} of the Euclidean four-point function, while by construction converges in the unit disc $|z|<1$, is not obviously convergent under the analytic continuation. In \cite{Maldacena:2015iua} (see also \cite{Fitzpatrick:2016mjq}), it was shown that the Virasoro block expansions of general four-point functions converge under arbitrary analytic continuations. 

Let us demonstrate this for the analytic continuation from the Euclidean four-point function \eqref{eqn:EuclideanWWVV} to the OTO four-point function \eqref{OtO4-pt}. Consider the following change of variables to render the four-point function \eqref{eqn:EuclideanCorrelator}  single-valued,
\ie
\label{eqn:qzmap}
q(z)=e^{i\pi\tau(z)},~~~\tau(z)=i{K(1-z)\over K(z)},~~~K(z)={\pi\over 2}\; {}_2F_1\bigg(\frac{1}{2},\frac{1}{2},1\bigg|z\bigg),
\fe
where the variable $q$ is called the elliptic nome.  The Virasoro block ${\cal F}(h_w,h_v,h;z)$ has a natural expression in terms of $q$ \cite{Zamolodchikov1987}g
\ie
\label{eqn:pillowF}
{\cal F}(h_w,h_v,h;z) &=  (16\,q)^{h-\ca}\; z^{\ca-2h_w} \; ( 1-z)^{\ca-h_w-h_v}
\\
&\qquad\qquad\times\, \theta_3(q)^{{c-1\over 2}-8(h_w+h_v)}\; H(h_w,h_v,h;q),
\fe
where the function $H(h_w,h_v,h;q)$ admits a series expansion in $q^2$ with the leading term $H(h_w,h_v,h;0)=1$. Furthermore, in the limit $h\to\infty$, the function $H(h_w,h_v,h;q)$ goes as
\ie
\label{eqn:Hinf}
\lim_{h\to\infty}H(h_w,h_v,h;q) =1.
\fe
We have found it convenient to also introduce a shifted central charge $\ca$, 
\begin{align}
\ca \equiv \frac{c-1}{24}\,,
\label{eq:ccdef}
\end{align}
to declutter subsequent formulae. 

\begin{figure}[H]
\centering
\subfloat{
\includegraphics[width=.4\textwidth]{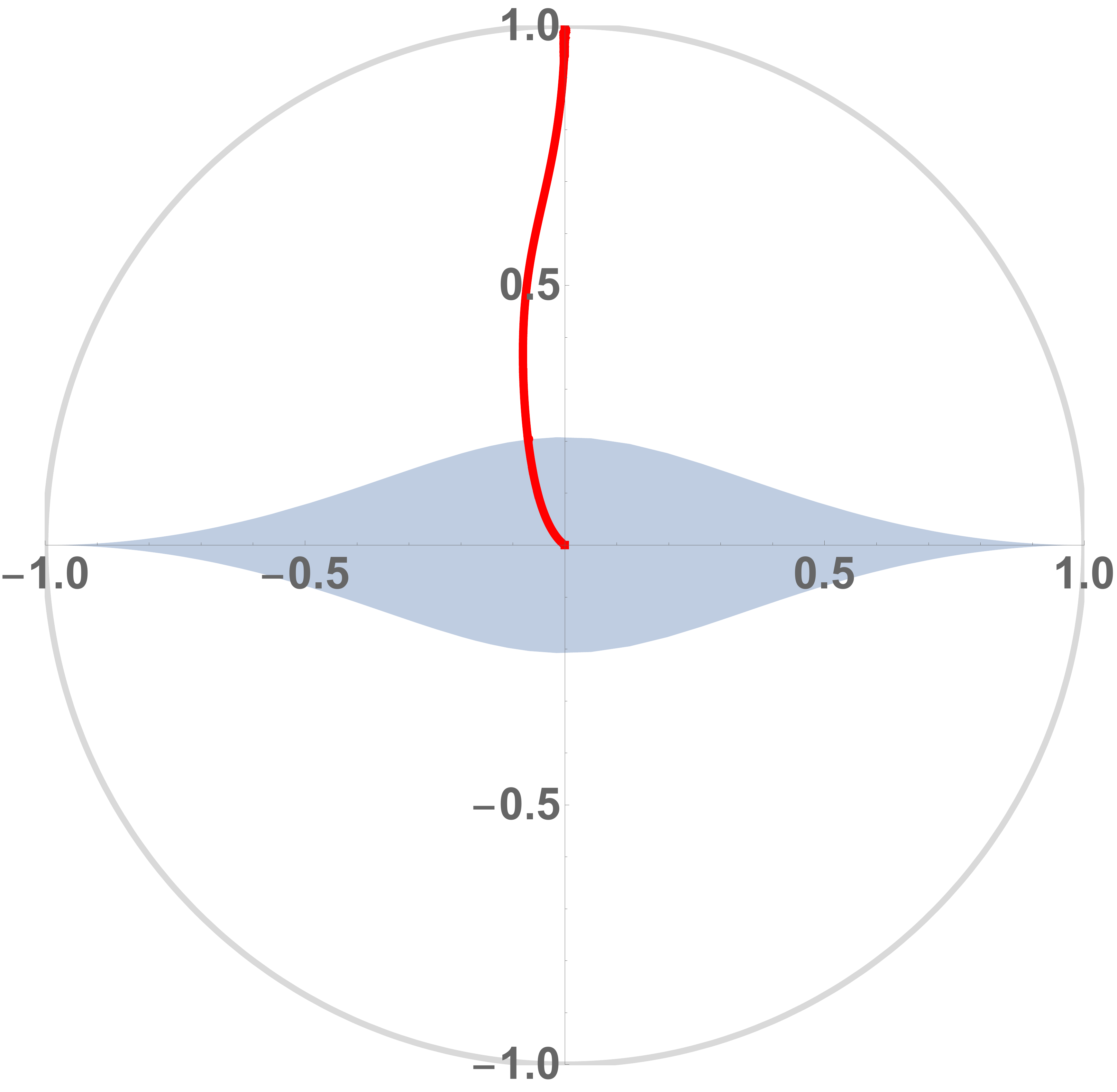}
}
\caption{The image of the map $q(z)$ given in Eq.~\eqref{eqn:qzmap} which takes us from the $\mathbb{C}$ parametrized by $z$ to the unit disc in $q$-space. The shaded domain is the entire $z$ plane and the red trajectory is a path  along which we analytically continue the Euclidean four-point function to the desired OTO correlator plotted here for 
$\epsilon_1=0,\,\epsilon_2=0.2,\,\epsilon_3=0.1,\,\epsilon_4=0.3,\,\beta=x=1$ and $t$ from 0 to $\infty$.}
\label{Fig:qPlane}
\end{figure}

The points $z=0$, 1, $\infty$ in the $z$-plane are mapped to the points $q=0$, $1$, $-1$ in the $q$-plane.  The entire $z$-plane is mapped to a compact region in the $q$-plane inside the unit circle $|q|\le 1$, which is shown as the blue shaded region in Fig.~\ref{Fig:qPlane}. The branch cut from $z=1$ to $z=\infty$ is mapped to the boundary of the blue shaded region. On the $q$-plane, the analytic continuation from the Euclidean four-point function \eqref{eqn:EuclideanWWVV} to the OTO four-point function \eqref{OtO4-pt} is along a path starting from a point in the blue shaded region to the white region inside the unit circle on the upper half plane. The point $t=\infty$ is mapped to $q=i$.
The convergence of the Virasoro block expansion \eqref{eqn:virasoroBlockDecomp}
along the interval $z\in[0,1]$ implies a bound on the OPE coefficients of operators of large dimensions
\ie\label{eqn:OPEbound}
16^{h+\bar h} \,  \big|C_{WW{\cal O}}\, C_{VV{\cal O}} \big| < 1\,.
\fe
More precisely, there exist a positive number $\Lambda$ such that the inequality \eqref{eqn:OPEbound} holds for all the operators ${\cal O}$ whose dimensions $h$ and $\bar h$ satisfy $h,\bar h > \Lambda$. This condition also implies the convergence of the Virasoro block expansion along the analytic continuation from the Euclidean four-point function \eqref{eqn:EuclideanWWVV} to the OTO four-point function \eqref{OtO4-pt}.

The function $H(q)$ has a nice interpretation as the Virasoro block of the four-point function on the pillow geometry $T^2/\bZ_2$ \cite{Maldacena:2015iua}. The $T^2$ can be viewed as an elliptic curve inside $\widehat\bC^2$ with the coordinates $(x,y)$ is defined by
\ie
y^2=x\, (z-x) \, (1-x),
\fe
which can be viewed as a double-cover of the Riemann sphere branched at 0, $z$, 1 and $\infty$, where $x$ is the coordinate of the base Riemann sphere $\widehat\bC$. The $\bZ_2$ action is generated by $y\to -y$. The map from the Riemann sphere $\widehat\bC$ to $T^2/\bZ_2$ is explicitly given by
\ie
x\mapsto u={1\over \theta_3(q)^2}\int^x_0{dw\over \sqrt{w(1-w)(z-w)}},
\fe
where $u$ is the coordinate of $T^2/\bZ_2$, which takes complex values with identification $u\sim u+2\pi\sim u+2\pi \tau$ and the $\mathbb{Z}_2$ identification is $u\sim -u$. The positions of the operators $(0,z,1,\infty)$ are mapped to $(0,\pi,\pi+\pi\tau,\pi \tau)$. Taking the conformal anomaly factor into account, the four-point function on $T^2/\bZ_2$ is given by
\ie
&\left<W(x=0)\, W(x=z)\, V(x=1)\, V(x=\infty)\right>_{\widehat\bC} 
\\
&= z^{{c\over 24}-2h_w}\, (1-z)^{{c\over 24}-h_w-h_v}\,\theta_3(q)^{{c\over 2}-8(h_w+h_v)}
\\
&\quad\times\left<W'(u=0)\, W'(u=\pi)\, V'(u=\pi+\pi\tau)\, V'(u=\pi\tau)\right>_{T^2/\bZ_2}.
\fe
The four-point function on $T^2/\bZ_2$ admits a decomposition into Virasoro blocks as\footnote{ We have used the identity
\ie
\theta_2(q)\theta_3(q)\theta_4(q) 
=(16\,q)^{1\over 4}\,\prod_{n=1}^\infty (1-q^{2n})^3.
\fe}
\ie
&\left<W'(0)\,W'(\pi)\,V'(\pi+\pi\tau)\,V'(\pi\tau)\right>_{T^2/\bZ_2}
\\
&=\sum_{a} C_{WW{\cal O}_a} \,C_{VV{\cal O}_a} {(16\,q)^{h_a-{c\over 24}}\; (16\,\bar q)^{\bar h_a-{c\over 24}}\over 
\prod_{n=1}^\infty (1-q^{2n})^{1\over 2}\, (1-\bar q^{2n})^{1\over 2}}\;H(h_w,h_v,h_a;q)
\; H(\bar h_w,\bar h_v,\bar h_a;\bar q).
\fe

\subsection{Semiclassical heavy-light limit}
\label{sec:semiclassical}

We have now assembled the ingredients to  review the analysis of \cite{Roberts:2014ifa} relating to the OTO four-point function in the semiclassical heavy-light limit. The semiclassical limit is defined by the $c\to \infty$ limit with ${h_w\over c}$, ${h_v\over c}$ and ${h\over c}$ fixed. The Virasoro block in such limit takes the exponential form as
\ie
{\cal F}(h_w,h_v,h;z)=\exp\left[-{c\over 6}\, f\left({h_w\over c},{h_v\over c}, \frac{h}{c} ;z\right)\right].
\fe
The heavy-light limit is defined on top of the semiclassical limit by further taking $\frac{h_v}{c}, \frac{h}{c} \to 0$ while holding 
$\frac{h_w}{c}$ fixed. The semiclassical Virasoro block in this limit was computed exactly in \cite{Hijano:2015rla}. Defining 
\ie
\A=\sqrt{1-{24\over c}\, h_w}.
\fe
one finds  the function $f$ from which we can extract
\ie\label{eqn:heavyLightBlock}
{\cal F}(h_w,h_v,h;z)
&=z^{2(h_v-h_w)}\left[{\A(1-z)^{{\A-1\over 2}}\over1-(1-z)^{\A}}\right]^{2h_v}\left[{4\left(1-(1-z)^{\A\over 2}\right)\over \A\left(1+(1-z)^{{\A\over 2}}\right)}\right]^h.
\fe

If we further take the small $z$ and small ${h_w\over c}$ limit of the heavy-light semiclassical block \eqref{eqn:heavyLightBlock} becomes (setting $\varepsilon \equiv {\cal O}(z^2,\tfrac{h_w^2}{c^2},\tfrac{z\,h_w}{ c}) $ for brevity)
\ie
z^{2(h_v-h_w)}\left({1\over z}+\varepsilon \right)^{2h_v}\left(z+\varepsilon \right)^h.
\fe

The exact formula for the heavy-light semiclassical block allows us to preform the analytic continuation $(1-z)\to e^{-2\pi i}(1-z)$,
\ie\label{eqn:ReggeBlock}
{\cal F}(h_w,h_v,h;z) 
& \overset{{{(1-z)\to e^{-2\pi i}(1-z)}}}{\xrightarrow{\hspace*{3cm}}} \\
&	z^{2(h_v-h_w)}
	\left[{\A \,e^{i\pi(1-\A)}\,(1-z)^{{\A-1\over 2}}\over1-e^{-2\pi i\A}\,(1-z)^{\A}}\right]^{2h_v}
	\left[{4\left(1-e^{-i\pi \A}(1-z)^{\A\over 2}\right)\over \A\left(1+e^{-i\pi\A}(1-z)^{{\A\over 2}}\right)}\right]^h.
\fe
In the small $z$ and small ${h_w\over c}$ limit, the analytic continued heavy-light semiclassical block  \eqref{eqn:ReggeBlock} behaves as
\ie
z^{2(h_v-h_w)}\left[{1\over z-{24\pi i \,h_w\over c}+\varepsilon }\right]^{2\,h_v}\left[16\over z-{24\pi i \,h_w\over c}+\varepsilon \right]^h.
\fe

Putting everything together, the OTO four-point function is given by
\ie\label{eqn:OOTO4-pt}
G(z,\bar z)&\approx \sum_{h_a} C_{VV{\cal O}_a}C_{WW{\cal O}_a}
\left({1\over 1-{24\pi i\, h_w\over c\,z}}\right)^{2\,h_v}\left(16\over z-{24\pi i \,h_w\over c}\right)^{h_a} \bar z^{\bar h_a},
\fe
where $\approx$ denotes the heavy-light semiclassical limit followed by the small $z$ and ${h_w\over c}$ limit.

\subsection{Chaos from the vacuum block}
\label{sec:vacbl}

We are interested in the intermediate time behavior of the OTO four-point function around the moment $t \sim t_*\gg x\ge 0$, where $t_*$ is the scrambling time
\ie
t_*={\beta\over 2\pi}\log c.
\fe
To examine this limit, let us first rewrite the Virasoro block expansion of the four-point function by isolating the contribution from the vacuum block as 
\ie\label{eqn:GExp}
&G(z,\bar z)=G_0(z,\bar z)\left[1+\sum_{(h,\bar h)\neq(0,0)} C_{VV{\cal O}}\,C_{WW{\cal O}}\;G_{h,\bar h}(z,\bar z)\right],
\\
&G_0(z,\bar z) \equiv z^{2h_w}\bar z^{2h_w}{\cal F}(0,z){\cal F}(0,\bar z),\quad
 G_{h,\bar h}(z,\bar z) \equiv { {\cal F}(h,z){\cal F}(\bar h,\bar z)\over  {\cal F}(0,z){\cal F}(0,\bar z)}.
\fe
When $t\gg x$, the $G_0(z,\bar z)$ and $G_{h,\bar h}(z,\bar z)$ become (setting $C \equiv  {24 \pi i\, h_w\over \epsilon_{12}\, \epsilon_{34}^*}$)
\ie\label{eqn:G0Gh}
G_0(x,t)&\approx\left({1\over 1+  C\, e^{{2\pi\over \beta}(t-x-t_*)}}\right)^{2h_v}, \quad
\\
G_{h,\bar h}(x,t) 
&\approx e^{{2\pi\over \B}\left[h t_*-\bar h(x+t)\right]}\left({16\over  e^{{2\pi\over \B}(x-t+t_*)}+C}\right)^h{ \left(- \epsilon_{12}\epsilon_{34}^*\right)^{\bar h}\over  \left(- \epsilon_{12}\epsilon_{34}^*\right)^{ h}},
\fe
 where we have used the formula of the cross ratios $z$ and $\bar z$ for $t\gg x$, 
\ie\label{eqn:zinxt}
&z  \approx  -e^{{2\pi\over \B}(x-t)}\epsilon_{12}\epsilon_{34}^*,
\quad
\bar z  \approx - e^{-{2\pi\over \B}(x+t)}\epsilon_{12}\epsilon_{34}^*,
\fe
where $\epsilon_{ab}=i(e^{{2\pi\over \beta}i\epsilon_a}-e^{{2\pi\over \beta}i\epsilon_b})$.

The vacuum block $G_0(x,t)$ was studied initially in \cite{Roberts:2014ifa} and then re-examined by \cite{Fitzpatrick:2016thx,Perlmutter:2016pkf}. We see that the vacuum block $G_0(x,t)$ starts to decay exponentially when the time $t$ approaches the scrambling time
\ie\label{eqn:vacuumDecay}
t-x\sim t_*.
\fe

The decay rate of the OTO correlator sets the Lyapunov exponent, which can be read off from \eqref{eqn:G0Gh} to be 
\ie
\lambda_L={2\pi\over \beta},
\fe 
which saturates the  universal bound on chaos \cite{Maldacena:2015waa}, viz.,
\begin{equation}
\lambda_L \leq \frac{2\pi}{\beta} \,. 
\label{eqn:chaosBound}
\end{equation}	
As argued in \cite{Roberts:2014ifa} the contribution from the vacuum block suffices to obtain this result.   However, in the late time limit, the pillow coordinate $q$ goes as
\ie
\label{eq:qchaoslimit}
q=i e^{-{\beta \pi\over 8 t}+{\cal O}(t^{-2})},
\fe
which approaches the boundary of the radius of convergence of OPE.  Hence, one expect the contributions from non-vacuum blocks become important when $t\sim t_*\gg 1$. The rest of the our discussion will be devoted to bounding the contribution from the non-vacuum blocks.

\section{Contribution from non-vacuum blocks}
\label{sec:nonvac}

We have seen that truncating the Euclidean correlator to the vacuum block contribution, and thence analytically continuing the result to the Lorentzian OTO domain appears to agree with the  classical gravity computation involving shock-wave states first carried out in \cite{Shenker:2013pqa}. One might a-priori view this as being due to the fact that the semiclassical  computation only cares about gravitational interactions, and thus should match with the results of the vacuum block. This is however misleading owing to the analytic continuation involved: it is unlikely to be the case that the vacuum block contribution in the Euclidean OPE channel is simply the graviton exchange in the semiclassical Lorentzian OTO channel. 

Our first task is to ask what is the contribution from the non-vacuum blocks and whether one can come up with constraints on the OPE coefficients to bound their contribution in a suitable way. Let us first start by noting some of the salient features of non-vacuum blocks.

\subsection{Timescale for non-vacuum block decay}
\label{sec:nonvacts}

The contribution $G_{h,\bar h}(x,t)$ from a non-vacuum block of dimension $(h,\bar h)$ contains an exponential factor
\ie
e^{{2\pi\over \B}\left[h \,t_*-\bar h\,(x+t)\, \right]},
\fe
which is small when
\ie\label{eqn:vacuumDominate}
t+x\;\gtrsim \;{h \over  \bar h}\,t_* \;\equiv\; t_s(h,\bar h).
\fe
 The conditions  \eqref{eqn:vacuumDecay} and \eqref{eqn:vacuumDominate} overlap for scalar primaries ($\ell=0$). Let us focus on the case $x=0$ and $h\ge\bar h$, and introduce the dimension, spin, and twist, respectively, of the intermediate operators, viz., 
\ie
\Delta=h+\bar h, \qquad \ell=h-\bar h, \qquad \tau=\Delta-|\ell| \,. 
\fe
We have  then
\ie 
\label{eqn:tshhb}
t_s(\Delta, \ell) = {h\over \bar h} \, t_*= t_* + {2\, \ell \over \tau}\, t_* \,,
\fe
For primary operators with finite twist and large spin, the inequality \eqref{eqn:vacuumDominate} becomes 
\ie\label{eqn:vacuumDominateApprox}
t\gtrsim {2\,\ell\over \tau} \,t_*,
\fe
which has no overlap with \eqref{eqn:vacuumDecay}.

This na\"ive estimate should also be supplemented to include the contribution from the OPE coefficients. As discussed in the previous section, the OPE coefficients of operators of large dimensions satisfy the bound \eqref{eqn:OPEbound}. Assuming the bound is saturated, the term $|C_{VV{\cal O}}\, C_{WW{\cal O}}|\; G_{h,\bar h}(z,\bar z)$ in the Virasoro block decomposition 
\eqref{eqn:GExp} is small when 
\ie
t+x\gtrsim {h \over \bar h}\,t_* - {\beta\over 2\pi}\left(   {h+\bar h\over \bar h}\log 16\right).
\fe
Let us again assume $x=0$, large positive spin and finite twist. We have
\ie\label{eqn:vacuumDominateCond}
t \gtrsim {2\,\ell\over \tau}\left(t_*- {\beta\over 2\pi}\log 16\right) .
\fe
Since the scrambling time $t_*={\beta\over 2\pi}\log c$ is much larger than ${\beta\over 2\pi}\log 16$ in the large $c$ limit,  the inequality \eqref{eqn:vacuumDominateApprox} remains unaffected. So all told, we see that if we have operators of large spin and finite twist, then the time scales for vacuum block dominance get pushed away from the scrambling time to \eqref{eqn:tshhb}. This was already noted in \cite{Roberts:2014ifa} (for the global blocks) and revisited recently in \cite{Liu:2018iki,Hampapura:2018otw}.

It was shown in \cite{Collier:2016cls} that, for unitary (compact or non-compact) 2d CFTs with $c>1$ and an SL(2) invariant normalizable vacuum, there are infinitely many large spin primaries whose twists accumulate to ${c-1\over 12}$. By this result, for any $t$ there are always infinitely many operators
with large enough ${\ell\over \tau}$ that violates the inequality \eqref{eqn:vacuumDominateApprox}. Hence, the contribution of them to the four-point function are large compared to the contribution from the vacuum Virasoro block. However, one should notice that in deriving the inequality \eqref{eqn:vacuumDominate}, we have used the formulas \eqref{eqn:heavyLightBlock} for the Virasoro blocks, which is only valid when $c\gg h\ge 0$ in the large $c$ limit. In  \S\ref{sec:num}, we compute the Virasoro blocks in the $q$ expansion numerically in high powers, and demonstrate that the contributions from large spin and low twist blocks individually are larger than the contribution from the vacuum block.

\subsection{Numerical results}
\label{sec:num}

We can explicitly check the contributions of the non-vacuum blocks by exploiting the representation of the Virasoro blocks in the pillow coordinate  \cite{Zamolodchikov1987,Zamolodchikov:1995aa,Maldacena:2015iua} reviewed in \S\ref{sec:pillow}. The explicit parameterization is given in \eqref{eqn:pillowF}.  The key fact we need is that  function $H(q)$ satisfies $H(0)=1$ and admits an expansion in $q^2$, where the expansion coefficients can be computed very effectively by Zamolodchikov's recursion relation \cite{Zamolodchikov:1985ie,Zamolodchikov1987,Zamolodchikov:1995aa}. Using the \texttt{C++} implementation in \cite{Chen:2017yze}, we compute the $q$-expansion up to ${\cal O}(q^{3000})$.\footnote{ As we shall review later \cite{Kusuki:2017upd, Kusuki:2018wcv, Kusuki:2018nms, Kusuki:2018wpa}  have made some useful progress extracting asymptotic features of the series coefficients from an explicit numerical solution of the recursion.}

The representation \eqref{eqn:pillowF} of the Virasoro blocks is particularly useful when we study the four-point function on different sheets. Let us consider the analytic continuation $(1-z)\to e^{-2\pi i}(1-z)$. We use the standard $z \leftrightarrow (1-z)$ formula of hypergeometric function to simplify the function $K(z)$ introduced in \eqref{eqn:qzmap}. We have
\ie
K(z)=-{1\over \pi}K(1-z)\log(1-z) + f(1-z),
\fe
where $f(z)$ is an analytic function in the neighborhood of $z=0$. Hence, we have
\ie
K(z)\to K(z) + 2 i \,K(1-z),\quad\tau(z)\to \widetilde\tau(z)= {\tau(z)\over 1+2\tau(z)},\quad\widetilde q\equiv e^{\pi i\widetilde \tau(z)}.
\fe
The Virasoro blocks  after analytic continuation are given by the formula
\ie\label{eqn:VBafterA}
G_0(z,\bar z) G_{h,\bar h}(z,\bar z) &=
\; e^{-2\pi i (\ca-h_w-h_v)}  \\
& \quad \times(16\,\widetilde q)^{h-\ca}\; z^{\ca-2h_v+2h_w} \;( 1-z)^{\ca-h_w-h_v}\;\theta_3(\widetilde q)^{{c-1\over 2}-8(h_w+h_v)}\; H(h_w,h_v,h;\widetilde q)
\\
&\quad\times (16\,\bar q)^{\bar{h}-\ca}\; \bar z^{\ca-2 \bar{h}_v+2 \bar{h}_w} \; ( 1-\bar z)^{\ca-\bar{h}_w-\bar{h}_v} \;\theta_3(\bar q)^{{c-1\over 2}-8(\bar{h}_w+\bar{h}_v)}\; H(\bar{h}_w,\bar{h}_v,\bar h;\bar q).
\fe

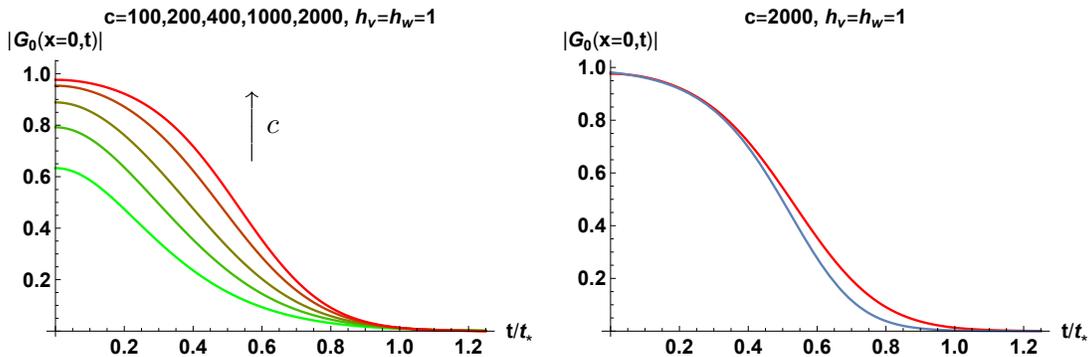
\begin{figure}[H]
\centering
\subfloat{
\includegraphics[width=.45\textwidth]{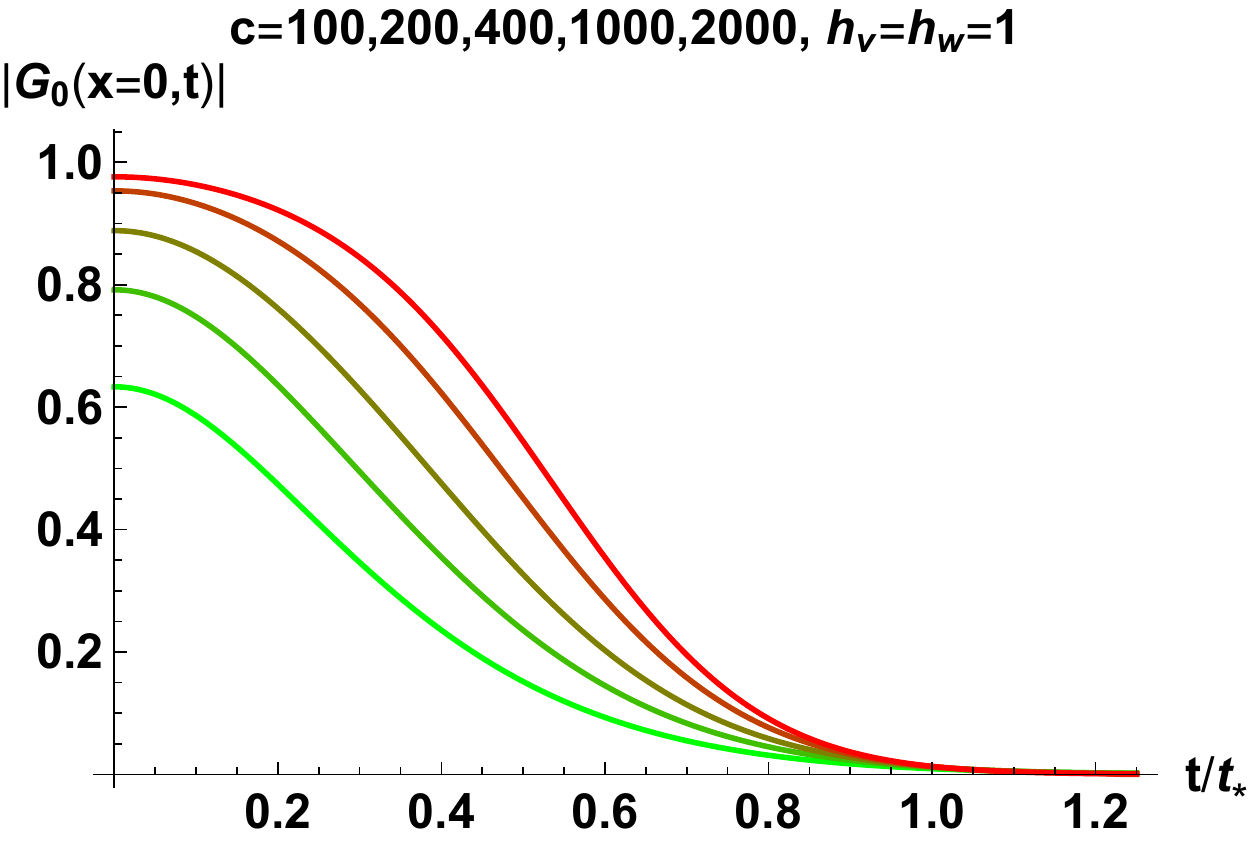}
\begin{picture}(0,0)
\setlength{\unitlength}{1cm}
\put (-4,3) {$\bigg\uparrow\,c$}
\end{picture}
}
\subfloat{
\includegraphics[width=.45\textwidth]{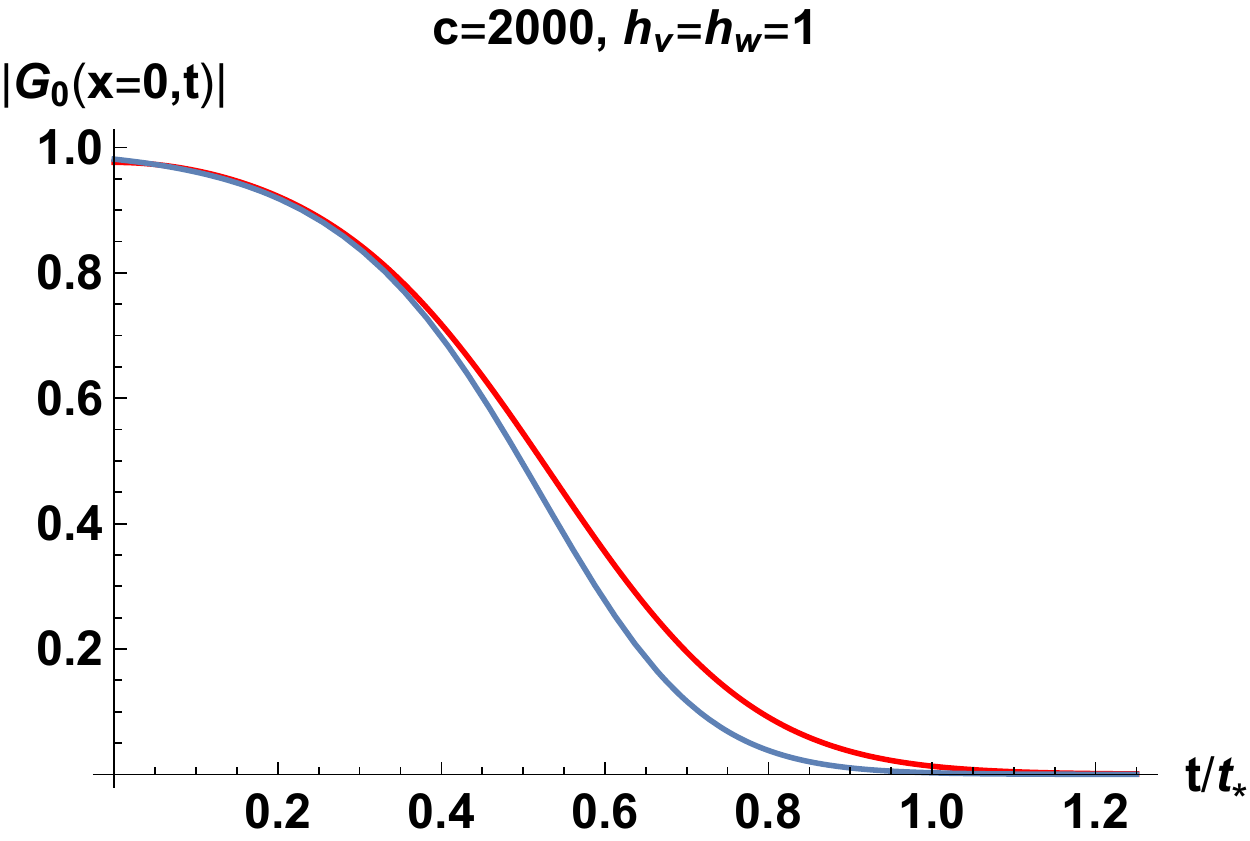}
}
\caption{The Virasoro vacuum blocks $G_0(x,t)$ in \eqref{eqn:GExp} at $x=0$ as a function of $t$ 
with $\beta=1$ and $(\epsilon_1,\epsilon_2,\epsilon_3,\epsilon_4)=(-{1\over 4},{1\over 4},0,{1\over 2})$. {\bf Left:} Increasing central charges $c$ are shown from green to red. {\bf Right:} A comparison between the numerically computed vacuum block \eqref{eqn:VBafterA} (shown in red) and the semiclassical heavy-light vacuum block \eqref{eqn:G0Gh} (shown in blue).}
\label{Fig:Vacuum}
\end{figure}
Armed with these results we can compute the Virasoro blocks numerically. We have considered central charge $c$ ranging between $100 - 2000$ and external operator's dimensions $h_v=h_w$ of order 1. At $t\ll t_*={\beta\over 2\pi}\log c$, the Virasoro blocks are approximated by the global blocks. As shown on the left of Fig.~\ref{Fig:Vacuum} (for $c=2000$), the vacuum Virasoro block is close to 1 at $t=0$ and remains a constant for small $t$.  When $t$ becomes of the order $t_*$, the vacuum Virasoro block decreases to zero. A comparison between the numerically computed vacuum block \eqref{eqn:VBafterA} and the semiclassical heavy-light vacuum block \eqref{eqn:G0Gh} is shown on the right of Fig.~\ref{Fig:Vacuum}. We can see that even though our configuration with $h_w=h_v=1$ is not in the heavy-light limit, the formula \eqref{eqn:G0Gh} of the semiclassical heavy-light vacuum block is still a good approximation.

\begin{figure}[H]
\centering
\subfloat{
\includegraphics[width=.47\textwidth]{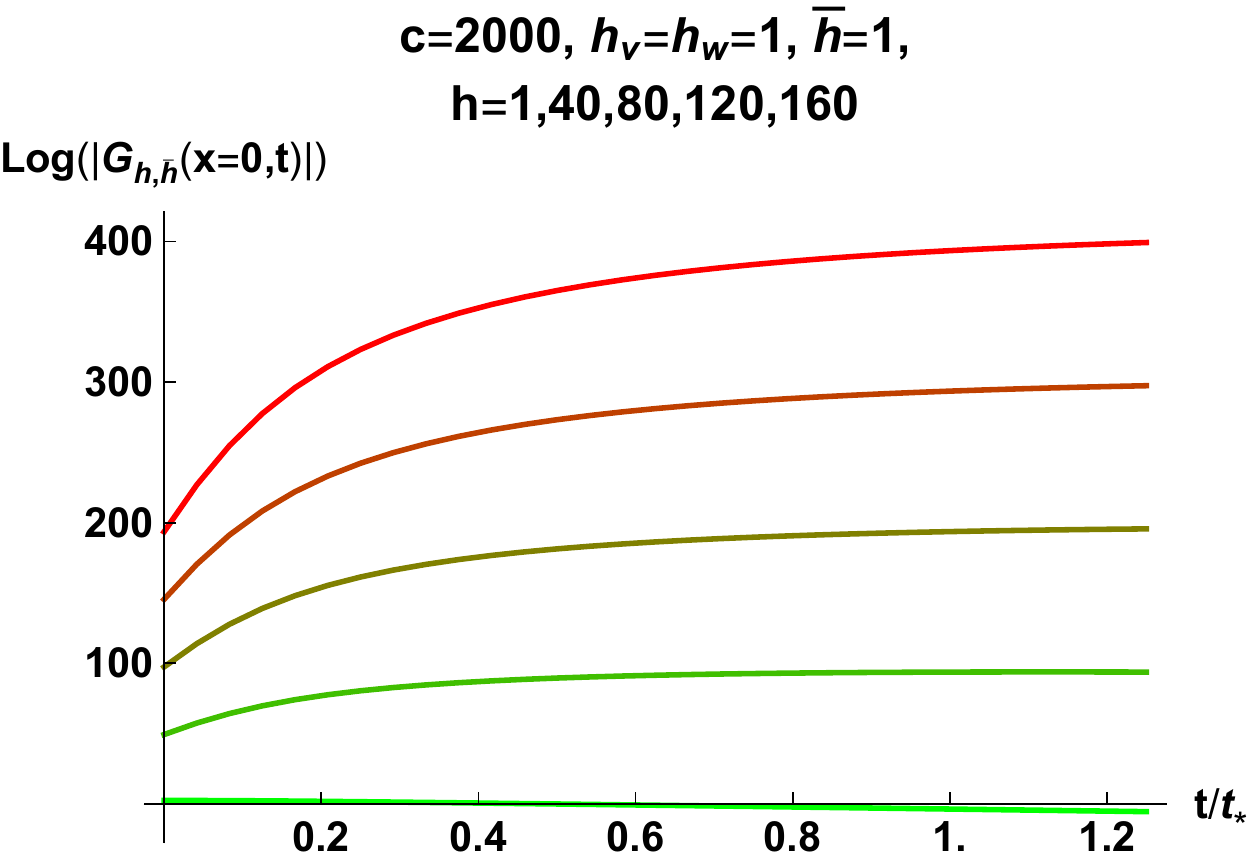}
}
\quad
\subfloat{
\includegraphics[width=.47\textwidth]{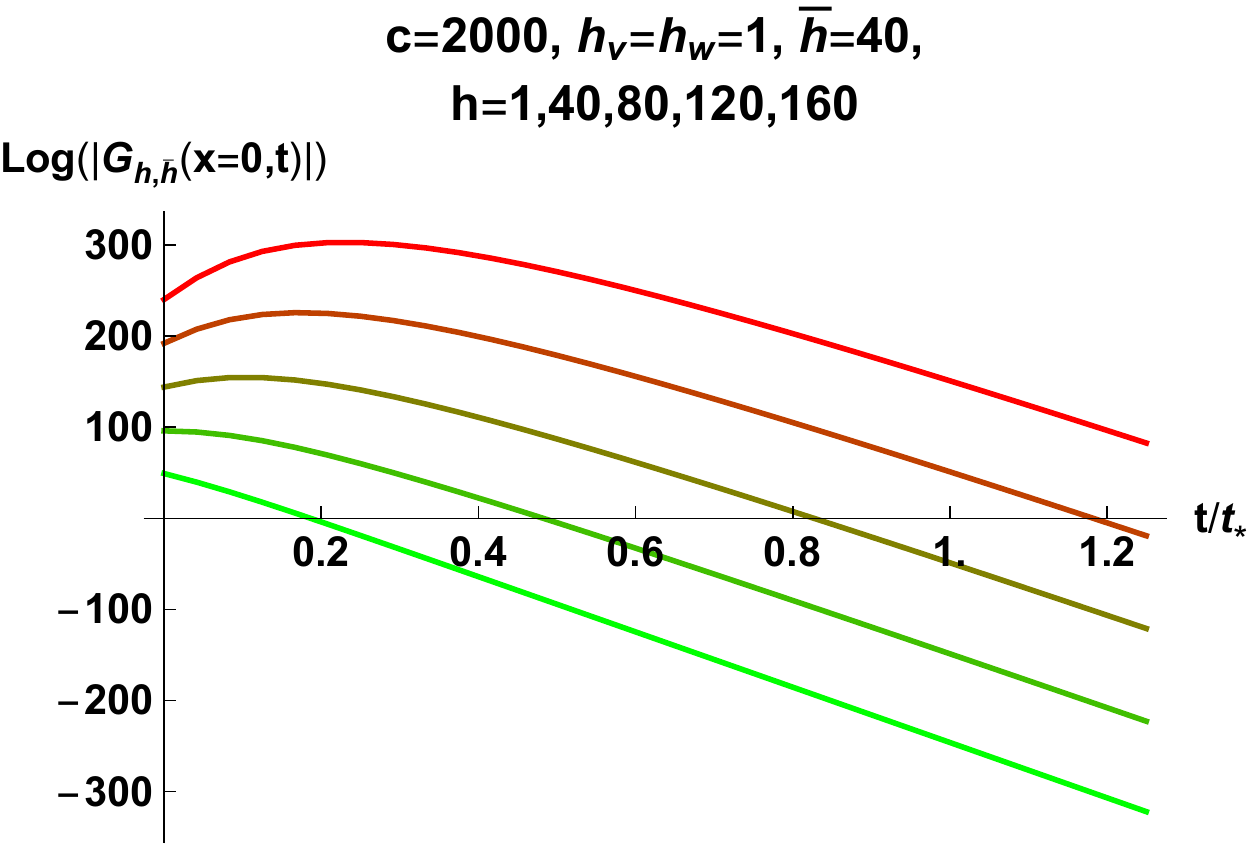}
}
\\
\subfloat{
\includegraphics[width=.47\textwidth]{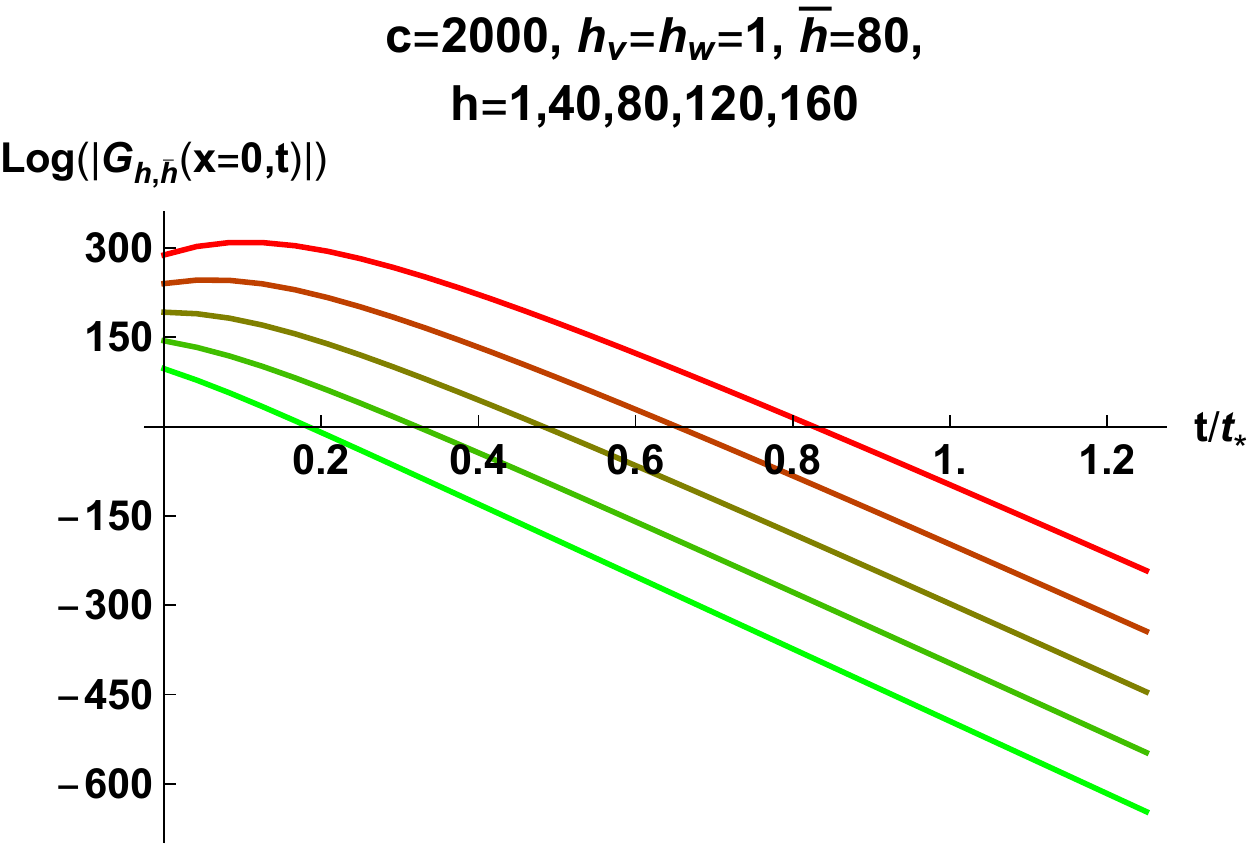}
}
\quad
\subfloat{
\includegraphics[width=.47\textwidth]{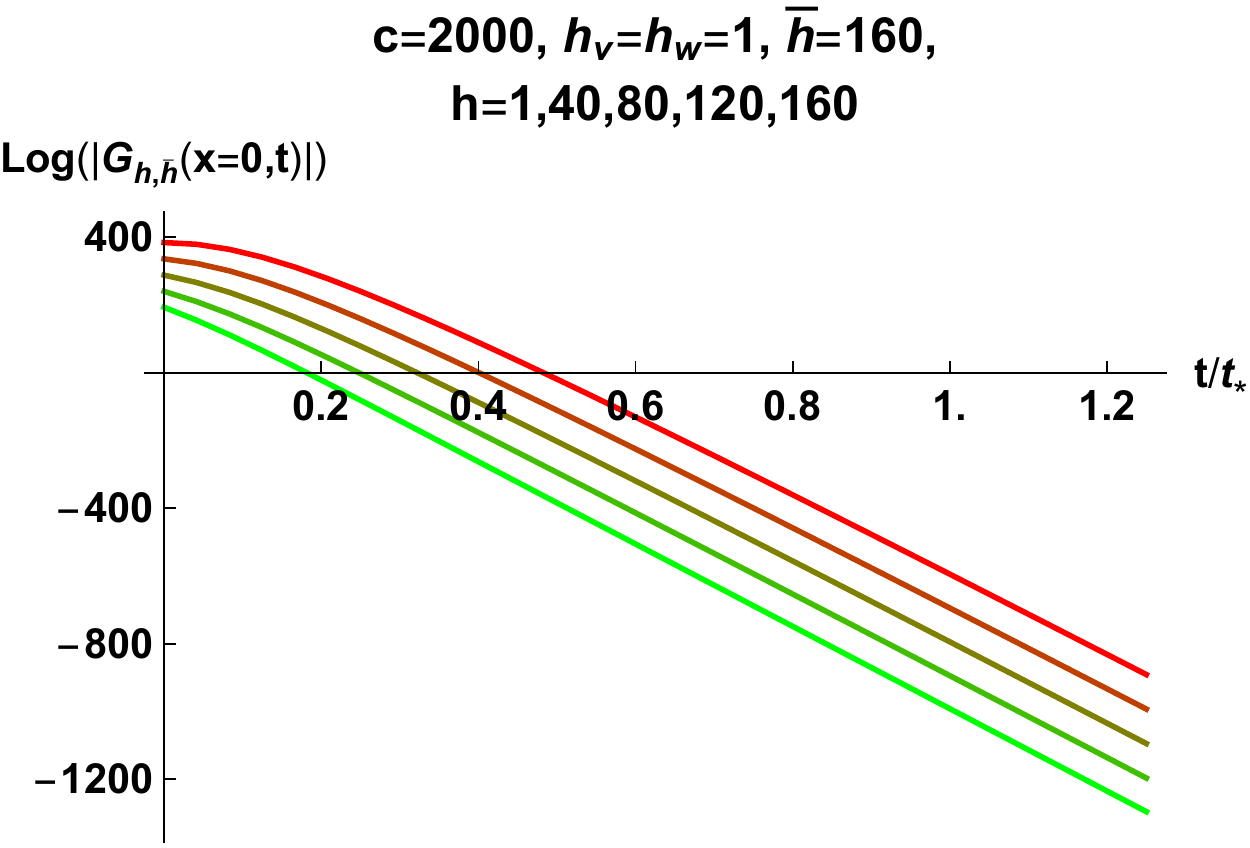}
}
\caption{ The logarithm of the ratio of the non-vacuum and vacuum blocks $G_{h,\bar h}(x,t)$ in \eqref{eqn:GExp} at $x=0$ as a function of $t$ with $\beta=1$ and $(\epsilon_1,\epsilon_2,\epsilon_3,\epsilon_4)=(0,{1\over 2},-{1\over 4},{1\over 4})$. Increasing dimensions $h$ of the exchanged operators are shown from green to red.}
\label{Fig:Ratios}
\end{figure}

The non-vacuum blocks are exponentially larger than the vacuum block in a range of time $0\le t\le t_s$, and the time $t_s$ increases when the spin $\ell=|h-\bar h|$ of the non-vacuum blocks increases, precisely as predicted from the analytic arguments in \S\ref{sec:nonvacts}. This is depicted in Fig.~\ref{Fig:Ratios} where we plot the ratio of the contribution from the non-vacuum block to the vacuum block for various choices of dimensions. 

\section{Bounds on non-vacuum contributions}
\label{sec:estimates}

We have see that the non-vacuum blocks have non-trivial contributions around the scrambling time. We might wonder if we can bound their combined effect  to remain sub-dominant. This requires on the one hand, getting a  handle on the OPE coefficients, and on the other  hand a more accurate  estimate of the block themselves, prior to analytic continuation. We will undertake this in a couple of steps. In \S\ref{sec:hspin}, we illustrate features that are transparent if we have operators with zero twist contributing to the OPE, as exemplified by theories with higher spin conserved charges. In \S\ref{sec:naive}, we proceed with a very na\"ive resummation over Virasoro blocks, and show that it gives $\lambda_L=\frac{2\pi}{\beta}$. Subsequently in the remaining subsections, we will try to refine our estimates, by isolating different domains in the conformal weights of the intermediate operator being exchanged.

\subsection{Higher spin theories}
\label{sec:hspin}

Consider a CFT with higher spin conserved currents in its primary operator spectrum. The higher spin currents have zero twist and finite spin. Hence, if they contribute to the $\vev{VVWW}$ four-point function, by the condition the condition \eqref{eqn:vacuumDominateApprox}, the Virasoro blocks associated to the higher spin currents would always dominate over the Virasoro vacuum block.

 One can instead study the vacuum block associated to the higher spin algebra generated by the higher spin conserved currents. It was shown in \cite{Perlmutter:2016pkf} that for a theory with higher spin conserved currents of bounded spin $s\le N$, the vacuum block of the higher spin algebra still exhibits a decaying behavior after the scrambling time $t_*$, but with the Lyapunov exponent is given by
\ie
\lambda_L={2\pi\over \beta}(N-1),
\fe
which violates the bound \eqref{eqn:chaosBound} for $N>2$. On the other hand, for a theory with higher spin conserved currents of unbounded spin, the vacuum block of the higher spin algebra does not decay in late time \cite{Perlmutter:2016pkf}. In other words, the Lyapunov exponent is zero.\footnote{ The vanishing of the Lyapunov exponent holds for theories with higher spin symmetries which nevertheless have a sparse spectrum as noted in \cite{Perlmutter:2016pkf} and reconfirmed in \cite{Belin:2017jli} for permutation orbifolds.}

The vacuum block of the higher spin algebra can be decomposed as a sum over Virasoro blocks weighted by OPE coefficients. In the above two cases, we see that different sums over the Virasoro blocks can lead to very different Lyapunov exponents.

\subsection{A na\"ive resummation}
\label{sec:naive}

Let us first preform a very simple-minded resummation over the Virasoro blocks, by making three drastic assumptions:
\begin{itemize}
\item First, we assume that CFT has a gap $(h_{\text{gap}},\bar h_{\text{gap}})$, and above the gap is an integer spectrum on 
$h$ and $\bar h$.\footnote{ When $h_{\rm gap}=h_v+h_w$ and $\bar h_{\rm gap}=\bar h_v+\bar h_w$, the spectrum coincides with the Regge trajectories \cite{Komargodski:2012ek,Fitzpatrick:2012yx,Fitzpatrick:2014vua,Collier:2018exn}. However, note that the OPE coefficients in the second assumption are not the same as the OPE coefficients in the (Virasoro) mean field theory.}
\item Second, we declare that  the  OPE coefficients saturate the bound \eqref{eqn:OPEbound}
\ie
C_{VV{\cal O}}\, C_{WW{\cal O}}=16^{-h-\bar h}.
\fe
\item Third, we use the explicit formulae \eqref{eqn:OOTO4-pt} and \eqref{eqn:G0Gh} for the Virasoro blocks in the semiclassical heavy-light limit.
\end{itemize}
Note that in typical CFTs the second assumption is only valid when $h,\bar h\gg c$; the third assumption however holds  when $h,\bar h\ll c$. Despite them not having a common regime of overlap, we will nevertheless proceed with these two estimates, if only to outline, in an uncontrolled approximation, a scenario where the resummation `works'.  

Armed with these we can resum the Virasoro block expansion \eqref{eqn:OOTO4-pt} quite easily, resulting in
\ie
G(z,\bar z)&\approx G_0(x,t)\left[1+\sum_{n,\bar n=0}^\infty 16^{-h_{\text{gap}}-\bar h_{\text{gap}}-n-\bar n}G_{h_{\text{gap}}+n,\bar h_{\text{gap}}+n}(x,t)\right],
\\
&=\left({1\over 1-{24\pi i \,h_w\over c\, z}}\right)^{2h_v}\left[1+{\left( z-{24\pi i \,h_w\over c}\right)^{1-h_{\text{gap}}}\over z-{24\pi i \,h_w\over c}-1}{16^{1-\bar h_{\text{gap}}}\bar z^{\bar h_{\text{gap}}}\over 16-\bar z}\right]
\\
&\approx\left({1\over 1-{24\pi i \,h_w\over c\, z}}\right)^{2h_v}-\left({1\over 1-{24\pi i \,h_w\over c\, z}}\right)^{2h_v+h_{\text{gap}}-1}z^{1-h_{\text{gap}}}\left(\bar z\over 16\right)^{\bar h_{\text{gap}}},
\fe
which can be equivalently written as (again $C = {2\pi i \,h_w\over \epsilon_{12}\epsilon_{34}^*} $)
\ie\label{eqn:naive_resum}
G(z,\bar z)\approx&\left({1\over 1+12\, C\, e^{{2\pi\over \beta}(t-x-t_*)}}\right)^{2h_v}
\\
&+{\epsilon_{12}\epsilon_{34}^*\over 16^{h_{\text{gap}}}}e^{{2\pi\over \beta}\left[(1-\Delta_{\text{gap}})x+(\ell_{\text{gap}}-1)t\right]}\left({1\over 1+12\, C\, e^{{2\pi\over \beta}(t-x-t_*)}}\right)^{2h_v+h_{\text{gap}}-1},
\fe
where $\Delta_{\text{gap}} = h_{\text{gap}}+\bar h_{\text{gap}}$ and $\ell_\text{gap}=h_{\text{gap}}-\bar h_{\text{gap}}$. The late time behavior is determined by the spin $\ell_\text{gap}$ of the lightest non-vacuum Virasoro primary operator. If we further assume $\ell_{\text{gap}}=0$ (we note that even if the lightest operator is spinless, the spectrum includes operators of arbitrary spin), then the second term is always smaller than the first term, and the result of the resummation just reduces back to the vacuum contribution.

This is of course, rather simple-minded and the constraints we imposed on the spectrum and OPE coefficients are too unrealistic. It however, does raise the possibility that in a class of theories one might indeed recover the physics associated with just the Virasoro vacuum block. The main question of course, is whether it is possible to show either generically or in certain specific family of CFTs that the resummation reaffirms the vacuum block dominance. We will try to address this question by trying to bound various limiting situations in the remainder of this section.

\subsection{Resumming heavy operators ($h,\,\bar h\gg c$)}
\label{sec:largehhb}

The preceding discussions have demonstrated how precise resummation of the conformal block decomposition can lead to qualitatively different behaviors in the chaos regime for $t\approx t_*$. Let us review this in a language that is best suited to the analysis of OPE convergence, viz., the pillow frame discussed in  \S\ref{sec:pillow}.

The chaos limit we seek is the regime  $z, \bar z \to 0$ with $z$ on the second sheet in the standard cross-ratio variables. This translates to the limit $q \to i$, i.e., $q$ is approaching the boundary of
the region where the $VV$ OPE converges in the $q$ coordinates, $|q| <1$. We have illustrated a particular trajectory along which one might carry out the desired analytic continuation in Fig.~\ref{Fig:qPlane}. As remarked above, we will now try to qualitatively assess the relative importance of the contributions of various subsets of the operators in the block decomposition.

To start, we first consider the contribution of  the very heavy intermediate operators, for simplicity in a correlator of identical operators $W=V$. We will take these to be primaries whose conformal dimensions are larger than by some arbitrarily chosen threshold $h_0\gg c$, over the value $ \frac{c-1}{24}$, viz., 
\begin{equation}
h \geq h_0 + \ca \,, \quad \bar h \geq \bar h_0 + \ca\,,   
\end{equation}	

The contribution of the such intermediate operators in the OPE expansion can then be written as
\begin{align}
	\label{eq:Ggtr}
  G_>(z,\bar z) ={}& \sum_{\substack{h>h_0+\ca, \\\bar h> \bar h_0+\ca}} \,
  				C_{VV}(h,\bar h)^2 \; {\cal F}_h(z) {\cal F}_{\bar h}(\bar z) \nonumber \\
  ={}& \int_{(h_0+\ca,\bar h_0+\ca)}^\infty dh\, d \bar h \; K(h,\bar h)  \; {\cal F}_h(z) {\cal F}_{\bar h} (\bar z)\, .
\end{align}
We have simplified the sum by representing it as an integral over
intermediate states, using the density of OPE coefficients
\begin{align}
	\label{eq:Kdef}
	K(h,\bar h) = \sum_{\mathcal{O}_{h',\bar h'}} C_{VV}(h', 
	\bar h')^2 \delta(h-h') \delta(\bar h - \bar h')\, .	
\end{align}
To resum \eqref{eq:Ggtr}, we need approximations for the density of OPE coefficients
$K(h, \bar h)$ and the Virasoro blocks
${\cal F}_h(z)$, for $h, \bar h \gg c$. We use the
following universal features:
\begin{enumerate}
\item For the conformal blocks we use the expression
  \eqref{eqn:pillowF} in terms of the pillow coordinate $q(z)$. It is
  worth noting that that $H_h(q)\to 1$ as $h\to \infty$,
  \eqref{eqn:Hinf}, so the full block in this limit is given by the
  universal prefactor
  \begin{align}
    {\cal F}(z) = (16q)^{h-\ca} \; z^{\ca-2h_v} \, (1-z)^{\ca-h_v-h_w} \, \theta_3(q)^{12\,\ca-8(h_v+h_w)} \, .
    \label{eq:Finf}
  \end{align}
  We emphasize that the only $h$ dependence in the blocks in this regime is the exponential suppression $q^h$.
\item The density of OPE coefficients $K(h,\bar h)$ for asymptotically
  heavy intermediate states can be determined by Cardy-like arguments,
  as shown in \cite{Das:2017cnv}. In particular,
  using the $q$ coordinates, the crossing equation translates into a
  statement about the modular behavior of the Virasoro block
  decomposition under $\tau \to {-} \frac{1}{\tau}$. Defining
  effective temperatures by $\tau(z) = \frac{i \beta(z)}{2\pi}$ and
  $\bar \tau(\bar z) = {-} \frac{i \bar \beta(\bar z)}{2\pi}$, vacuum
  dominance in the $\beta, \bar \beta \to \infty$ limit,
  i.e., $z, \bar z \to 0$, and the asymptotic behavior of the blocks
  \eqref{eq:Finf} leads to the following universal behavior of the OPE
  coefficients for heavy intermediate states $h, \bar h \gg c$:
  \begin{align}
    \label{eq:Kinf}
    K(h,\bar h) \equiv{}& \kappa(h) \kappa(\bar h)\, ,
  \end{align}
  where
  \begin{align}
    \label{eq:kappadef}
    \kappa(h) ={}& 16^{{-}h} e^{2\pi \sqrt{\ca(h-\ca)}}
  \end{align}
	Here we have worked to order $\sqrt{h}$ in the exponent for simplicity; subleading corrections are readily obtained.  We note that this expression is for the density of OPE coefficients, so that the average coefficients for large $h, \bar h$ can be found using the Cardy density of states:
	\begin{align}
		\overline{C^2(h, \bar h)} = \frac{K(h,\bar h)}{\rho(h, \bar h)} \sim 16^{-h-\bar h} e^{{-}2\pi \sqrt{\ca} (\sqrt{h-\ca} + \sqrt{\bar h - \ca})}\, .	
	\end{align}
	Thus we are essentially including subleading corrections to the bound $C^2 \sim 16^{-h-\bar h}$ at large $h, \bar h$. 
\end{enumerate}

Combining these two ingredients, we see that the 
resummation \eqref{eq:Ggtr} holomorphically factorizes 
and the resummation is governed by the following integral 
(we recall that $|q| < 1$):
\begin{align}
  \label{eq:Idef}
	I(h_0;q)\equiv{}& \int_{h_0+\ca}^\infty d h \, 
	(16 q)^{h-\ca} \kappa(h) \sim \int_{h_0}^\infty d \tilde h
	\, e^{{-} \gamma \tilde h + 2\pi 
	\sqrt{\ca \tilde h}}\, , 
\end{align}
where we have only kept terms to the order we are working, shifted the integration variable to $\tilde h= h- \ca$, and defined $\gamma = {-} \log q$. This integral can be evaluated to yield 
\begin{align}
	I(h_0;q) \sim{}&  \sqrt{\frac{\ca}{\pi \gamma}} \frac{\partial}{\partial \ca}
	\left[e^{\pi^2 \ca/\gamma} \erfc \left( \sqrt{\gamma h_0} - \pi \sqrt{\tfrac{\ca}{\gamma}} \right) \right]
	  \nonumber \\
	\sim{}& \frac{q^{h_0} e^{2\pi \sqrt{\ca h_0}}}{\gamma}  \, .
\end{align}
Here $\erfc(z) = 1 - \erf(z)$ is the complementary error function, and in the second line we have kept only the leading term for large $h_0$, 
using the asymptotic approximation
\begin{align}
	\erfc(z) \sim \frac{e^{{-} z^2}}{\pi z} \left(1 + {\cal O}(z^{{-}1}) \right)\, ,
\end{align}
for $|z| \to \infty$.

With this result, we can approximate \eqref{eq:Ggtr} by
\begin{align}
	\label{eq:Ggtrapprox}
	G_> \sim{}&  \frac{q^{h_0} \bar q^{\bar h_0} }{\gamma \bar \gamma}\bar [z \bar z (1-z)(1-\bar z)]^{\ca - 2 h_v}
	 [\theta_3(q) \theta_3(\bar q)]^{12 \ca - 16 h_v}   \, .
\end{align}
We can now continue this contribution to the chaos regime using \eqref{eq:qchaoslimit} and 
$\bar q \to \frac{\bar z}{16}$ as $\bar z \to 0$. The behavior of $\theta_3(q)$ as $q \to i$
can be determined by the modular properties of $\theta_3$, which tells us
\begin{align}
	\theta_3(\tau) = [{-}i({-}2 \tau+1)]^{{-}\frac{1}{2}}
	\theta_3\left(\frac{\tau}{{-}2\tau+1} \right)\, .		
\end{align}
The limit \eqref{eq:qchaoslimit} corresponds to $\tau \sim \frac{1}{2} - \frac{i \pi}{4\log (z/16)}$ (with $z\to 0$) 
and therefore
\begin{align}
	\theta_3(q\sim i) \to \left[\frac{\pi}{2\log(z/16)} 
	\right]^{{-}\frac{1}{2}} 	\, ,	
\end{align}
where we have used $\theta_3(\tau \to i \infty) = 1$. Taking
the limits in \eqref{eq:Ggtrapprox}, we end up with
\begin{align}
  G_> \sim{}& {-}\frac{2i e^{{-}2\pi i (\ca - \Delta_v)}}{\pi \log(\bar z/16)} z^{\ca - \Delta_v} \bar z^{\bar h_0 - \Delta_v} \left[ \frac{\pi}{2 \log(z/16)} \right]^{4 \Delta_v - 6 \ca} \nonumber \\
	\equiv{}& P(x,t) e^{{-}\frac{2\pi}{\beta} [(\bar h_0 + \ca-2 \Delta_v )t + (\bar h_0 - \ca) x ]}
  \label{eq:Gheavy}
\end{align}
where in the last line we have grouped all of the prefactors,
which are sub-exponential in $t$, into $P(x,t)$, which is 
explicitly given by
\begin{align}
	P(t) ={}& {-} \frac{2i e^{{-}2\pi i (\ca - \Delta_v)}}{\pi \log(\bar z/16)} \left[ \frac{\pi}{2 \log(z/16)} \right]^{4 \Delta_v - 6 \ca} \nonumber \\
	={}&  \frac{2i e^{{-}2\pi i (\ca - \Delta_v)}}{\pi \left[ \frac{2 \pi}{\beta} (x+t) -4 \log 2 \right]}\left[ \frac{\pi}{\frac{4 \pi}{\beta} (x-t) - 8\log 2} \right]^{4 \Delta_v - 6 \ca}
\end{align}
One can check that $P(t \to \infty)$ simplifies to a 
constant.

The key takeaway of \eqref{eq:Gheavy} is the  manifest exponential  decay in $t$, demonstrating that the contribution of the very 
heavy operators is suppressed in the chaos regime.

\subsection{Resumming light operators ($h,\,\bar h\ll c$)}
\label{sec:smallhhb}

The very heavy intermediate states are well under control and give  a nicely decaying contribution. Let us now turn to the opposite regime where the operators being exchanged are light and estimate their contribution to the chaos correlator. 

The ingredients we will use are the `sparse' density of states (see \cite{Hartman:2014oaa}) and corresponding OPE coefficients determined in
\cite{Chang:2015qfa,Chang:2016ftb}, as well as the semiclassical conformal blocks used in \S~\ref{sec:semiclassical}. In 
\cite{Chang:2015qfa,Chang:2016ftb} it was demonstrated that requiring the vacuum block dominate the semiclassical conformal block expansion in the Euclidean regime $0<z<1$ constrains the `light' OPE coefficients $C_{VV}^2(h, \bar h)$ (smeared over the spectrum of operators near $h$, $\bar h$) to behave as 
\begin{align}\label{eqn:lightOPEbound}
  \frac{6}{c} \log \bigg(\rho(h, \bar h) C^2_{VV}(h, \bar h)\bigg) \lesssim{}& 
  	f\left( \frac{h_v}{c}, \frac{h}{c}; \frac{1}{2} \right)- f\left(\frac{h_v}{c}, \frac{h}{c}; 0\right) \nonumber \\
  &+ (\text{anti-holomorphic})\, ,
\end{align}
which we assume holds for $h, \bar h < m_v\, c$, for some reasonable choice of $m_v(h_v)$ (see below).  

These results, as indicated by the notation, rely on the operators $V$ and $W$ being identical, so we have assumed that we can take $C_{VV} \simeq C_{WW}$.  We expect that the results can be  generalized, and proceed with the estimate without further ado. 
We shall work in the limit $h_v \ll c$, using  the semiclassical blocks described above
(expanding $\alpha \approx 1 - 12\, \frac{h_v}{c}$, making use of our assumption that $h_v \simeq h_w$), 
in which case $m_v(h_v) \,c \approx \sqrt{2} \,h_v$,

Assuming the OPE coefficients saturate this bound in the
regime $h < \epsilon \,c\approx m_v\, c$ with $\epsilon\ll 1$, we can estimate the contribution of a Virasoro block of conformal dimension $(h,\bar h)$ to the correlator as
\ie\label{eqn:single_light_block}
&(z\bar z)^{2h_w}C^2_{VV}(h,\bar h) \; {\cal F}_h(z) {\cal F}_{\bar h}(\bar z)
\\
& = (z\bar z)^{2h_w}\exp\left\{\frac{c}{6}\left[ f\left(\frac{h_v}{c},\frac{h}{c};\frac{1}{2}\right) - 
  		f\left(\frac{h_v}{c},\frac{h}{c};0\right) - f\left(\frac{h_v}{c},\frac{h}{c};z\right)\right]  + (\text{anti-hol.})\right\}  
\\
&=G_0(z,\bar z)\left[g(z)g(\bar z)\right]^{\tau\over 2}g(z)^{{1\over 2}(|\ell| + \ell)} g(\bar z)^{{1\over 2}(|\ell| - \ell)}
\fe
where we introduced the spin $\ell = h-\bar h$ and twist $\tau = (h+\bar h) - |\ell| $, and the functions
\begin{align}
G_0(z,\bar z)&=   (z\bar z)^{2h_v} \left( \frac{\alpha (1-z)^{\frac{\alpha-1}{2}}}{1-(1-z)^\alpha}
       \right)^{2 h_v}\left( \frac{\alpha (1-\bar z)^{\frac{\alpha-1}{2}}}{1-(1-\bar z)^\alpha}
       \right)^{2 h_v},
       \\
g(z)& = \left(\frac{1+2^{-\frac{\alpha}{2}}}{1-2^{-\frac{\alpha}{2}}} \right) \left(
  \frac{1- (1-z)^{\frac{\alpha}{2}}}{1+(1-z)^{\frac{\alpha}{2}}}\right)\, .
\end{align}
We consider analytic continuation $(1-z)\to e^{-2\pi i}(1-z)$, and take the small $z$, $\bar z$ and ${h_w\over c}$ limit, we find (setting $\varepsilon \equiv {\cal O}(z^2,\tfrac{h_w^2}{c^2},\tfrac{z\,h_w}{ c}) $ for brevity)
\ie
&G_0(z,\bar z)=  (z\bar z)^{2h_v} \left({1\over z-{24\pi i \,h_w\over c}+\varepsilon }\right)^{2\,h_v}\left({1\over \bar z+\bar \varepsilon }\right)^{2\,h_v},
\\
&g(z)={4(3+2\sqrt{2})\over z-{24\pi i \,h_w\over c}+\varepsilon},\quad g(\bar z)={3+2\sqrt{2}\over 4}(\bar z+\bar \epsilon).
\fe
In this limit, we see that the behavior of \eqref{eqn:single_light_block} is determined only by the spin $\ell$ but not the twist $\tau$. This motivates us to sum over the contributions with spins in the range $\ell_{\rm gap}<\ell<\ell_{\rm max}\approx \epsilon \, c$ and a fixed twist $\tau<\epsilon \, c-|\ell|$,
\ie\label{eqn:sum_fix_twist}
G_{<}(z,\bar z,\tau)&= G_0(z,\bar z)\sum_{\ell=\ell_{\rm gap}}^{\ell_{\rm max}}\,\left(g(z)^{{1\over 2}(\tau +|\ell| + \ell)}g(\bar z)^{{1\over 2}(\tau+|\ell| - \ell)}+g(z)^{{1\over 2}(\tau +|\ell| - \ell)}g(\bar z)^{{1\over 2}(\tau+|\ell| + \ell)}\right)
\\
&\approx \left(3+2\sqrt{2} \right)^{\tau }\left({1\over 1-{24\pi i \,h_w\over cz} }\right)^{2\,h_v+{\tau\over 2}}\left(\bar z\over z\right)^{\tau\over 2}
\\
&\quad\times\left[z^{-\ell_{\rm max}}\left({4(3+2\sqrt{2})\over 1-{24\pi i \,h_w\over cz}}\right)^{\ell_{\rm max}}-z^{1-\ell_{\rm gap}}\left({4(3+2\sqrt{2})\over 1-{24\pi i \,h_w\over cz}}\right)^{\ell_{\rm gap}-1}\right].
\fe
This can be understood as summing over operators in a ``quantum Regge trajectory" introduced in \cite{Collier:2018exn}, which arises when decomposing a T-channel Virasoro vacuum block in terms of S-channel Virasoro blocks.\footnote{ The explicit formula of the quantum Regge trajectories is given in (1.8) of \cite{Collier:2018exn}. The spin of the operators generically are not integer; however, in the large $c$ limit, they become non-negative integers. However, note that the OPE coefficients \eqref{eqn:lightOPEbound} used in the sum \eqref{eqn:sum_fix_twist} are not the same as the OPE coefficients of the Virasoro mean field theory.}

Using the asymptotic expressions \eqref{eqn:zinxt} for $z$ and $\bar z$, we find that
\ie\label{eqn:light_sum_final}
&G_{<}(z,\bar z,\tau)\approx B\left({1\over 1+  C\, e^{{2\pi\over \beta}(t-x-t_*)}}\right)^{2\,h_v+{\tau\over 2}}e^{-{2\pi\over \beta}\tau x}
\\
&\quad\times\left[e^{{2\pi\over \beta}\ell_{\rm max}(t-x)}\left({A\over 1+  C\, e^{{2\pi\over \beta}(t-x-t_*)}}\right)^{\ell_{\rm max}}-e^{{2\pi\over \beta}(\ell_{\rm gap}-1)(t-x)}\left({A\over 1+  C\, e^{{2\pi\over \beta}(t-x-t_*)}}\right)^{\ell_{\rm gap}-1}\right],
\fe
where $A$, $B$, and $C$ are numerical factors
\ie
&A=-{4(3+2\sqrt{2})\over \epsilon_{12}\epsilon_{34}^*},\quad B=\left(3+2\sqrt{2} \right)^{\tau }\left(\epsilon_{12}^*\epsilon_{34}\over \epsilon_{12}\epsilon_{34}^*\right)^{\tau\over 2},\quad C={24\pi i h_w\over \epsilon_{12}\epsilon_{34}^*}.
\fe
We see that the leading term in the square bracket grows exponentially in late time $t\sim t_*\gg {2\pi\over \beta},x$, reflecting features we have seen earlier in the analysis. Clearly, the light operators by themselves would invalidate the conclusions inferred by truncating to the vacuum conformal block. One expects that the reconciliation of the growth with the chaos bound expectations reviewed in \S\ref{sec:review} will require cooperation of light and intermediate weight operators in the $VV$ OPE. More explicitly, the leading term depends on the scheme dependent spin cutoff $\ell_{\rm max}$, which with the dimension cutoff $\epsilon\, c$ sets an artificial boundary between the light and intermediate weight operators. Including the contribution from the intermediate weight operators is expected to remove the cutoff dependence of the correlator. A very plausible situation is that the leading term in \eqref{eqn:light_sum_final} would be canceled by such contribution. On the other hand, the subleading term in \eqref{eqn:light_sum_final} agrees nicely with our previous result \eqref{eqn:naive_resum} from a na\"ive resummation.  Therefore, in order for the non-vacuum block contribution to be subdominant than the vacuum block, our result suggests that for holographic CFTs the quantum Ragge trajectories should admit extension to spin $\ell_{\rm gap}=0$ or $1$.

\subsection{Resumming large spin operators ($h\gg c,\, \bar h<\mathfrak{c}$ or $\bar h\gg c,\, h<\mathfrak{c}$)}

In this section we will resum the asymptotic tails of the quantum
Regge trajectories introduced in \cite{Collier:2018exn}. As discussed
there, the presence of the vacuum block in the T-channel conformal
block decomposition can be used to extract S-channel OPE data using
the Virasoro fusion kernel. For sufficiently light external operators
(we shall again restrict to the case $h_v = h_w$ for simplicity), the
spectrum deduced in \cite{Collier:2018exn} holomorphically factorizes,
with the chiral spectrum is supported on a discrete set of weights
$h_m < \ca$ as well as continuous series for $h> \ca$ and similarly
for the antichiral part. The full operator spectrum is given by a
product of the holomorphic and antiholomorphic pieces, leading to four
regions in the $(h, \bar h)$ plane; this spectral data, extracted by
the T-channel vacuum block, was termed Virasoro mean field theory
(VMFT) in \cite{Collier:2018exn}. For unitary compact CFTs with $c>1$ and a positive lower bound on the twists of non-vacuum primaries, the spectrum and OPE coefficients of primary operators with $h\gg c$ or $\bar h\gg c$ universally approach those in the VMFT \cite{Collier:2018exn}. We have in fact already encountered a
portion of this spectrum, namely in section \S\ref{sec:largehhb}, where we
resummed the asymptotically heavy operators, $h, \bar h \gg c$. One
can check that the spectral density we used matches that extracted via
the fusion kernel in \cite{Kusuki:2018wpa,Collier:2018exn}.

Furthermore, resumming the discrete-discrete part of the VMFT spectrum
(in the large $c$ limit we are interested in) reproduces the exchange
of the vacuum operator in the T-channel, i.e.~reproduces
$\vert 1-z\vert^{-2\Delta_v}$, which is obviously non-singular in the
chaos regime. In fact, in the large $c$ limit this portion of the
spectrum reduces, to leading order, to that of the familiar
double-twist operators of global MFT spectrum.

In the rest of this section, we consider the asymptotically large spin
portions of the VMFT spectrum, which arise from combining a discrete
holomorphic operator with the continuous antiholomorphic spectrum (and
vice versa). 

We first consider the case with fixed $h$ and varying $\bar h$. That
is, we set $h = h_m \equiv 2h_v+ m +\delta h_m$ (here $\delta h_m$ is
an exact anomalous dimension due to summation of all multi-traces
built out of stress tensors) and integrate over $\bar h$ using the
antiholomorphic half of the spectral density we used in
\S\ref{sec:largehhb}. More concretely, as shown in
\cite{Collier:2018exn}, the spectral density obtained from the fusion
kernel factorizes into holomorphic and antiholomorphic pieces, with
the antiholomorphic half given by $\kappa(\bar h)$, defined in
\eqref{eq:kappadef} and the holomorphic half given by a residue of the
fusion kernel:
\begin{align}
  C_{VV}^2(h_m, \bar h) \sim {-} 2 \pi \kappa(\bar h) \Res\limits_{\alpha_s=\alpha_m} S_{\alpha_s 1} \, .
\end{align}
Here $S$ is the fusion kernel and $\alpha_m$ is defined via
$h_m = \alpha_m(Q-\alpha_m)$ (where $Q^2 = \frac{c-1}{6}$), which
rewrites a T-channel Virasoro block in terms of S-channel Virasoro
blocks; we refer the reader to \cite{Collier:2018exn} for its explicit
form (which we won't need in detail). The contribution of the
$\bar h \gg c$ portion of the $m$th quantum Regge trajectory is then
given by
\begin{align}
  G_{m,>} (z,\bar z) \equiv {\cal F}_{h_m}(z) \int_{\bar h_0 + \ca} d \bar h\, C_{VV}^2(h_m, \bar h) {\cal F}_{\bar h}(\bar z) \sim \left({-} 2\pi \Res_{\alpha_s=\alpha_m} S_{\alpha_s, 1} \right) {\cal F}_{h_m}(z) I(\bar h_0; \bar q)\, ,
\end{align}
where $I(\bar h_0; \bar q)$ is given by \eqref{eq:Idef}. We want to
continue this sum to the chaos regime. The antiholomorphic
continuation is straightforward, since
$I(\bar h_0, \bar q) \propto \bar q^{\bar h_0}$ and
$\bar q \sim \frac{\bar z}{16}$ in the chaos regime. For the
holomorphic half, we can use the behavior of the semiclassical blocks
in the chaos region described in \S\ref{sec:vacbl}. Combining the two
pieces, we find
\begin{align}
  G_{m,>}(z, \bar z) \propto e^{\frac{2 \pi}{\beta}(h_m t_* - \bar h_0 t)}\, ,
\end{align}
up to ratios of polynomials in $t$. Since by construction we have
taken $\bar h_0 \gg c$, we see that the tail ends of each quantum
Regge trajectory are in fact exponentially suppressed in the chaos
region.

Here we have assumed that the holomorphic part of the OPE
coefficients, i.e., $\Res S_{\alpha_s 1}$, does not have any
interesting features in the large $c$ limit. This can be ensured by
taking $h_v$ sufficiently small, i.e., scaling sufficiently weakly with
$c$. Explicit expressions for these residues can be found in
\cite{Collier:2018exn}. In this case, the leading order residues are
simply the (chiral half of) global MFT double-twist OPE
coefficients. We leave a more detailed analysis of perhaps more
interesting semiclassical limits to future work.

A similar story holds for the $\bar h$ fixed, $h > h_0 + \ca$
trajectories as well. Using the factorized form of the OPE
coefficients, one finds
\begin{align}
  G_{>,m}(z, \bar z) \sim \left({-} 2\pi \Res_{\bar \alpha_s=\bar \alpha_m} S_{\bar \alpha_s, 1} \right) {\cal F}_{\bar h_m}(\bar z) I(h_0; q)\, .
\end{align}
Using \eqref{eq:qchaoslimit}, the contribution of this sector is again
suppressed in the chaos limit, as the anti-holomorphic block is
exponentially suppressed and one has
\begin{align}
  G_{>,m}(z, \bar z) \propto e^{{-}\frac{2\pi}{\beta}\bar h_m t -\frac{ \pi\beta}{8\,t} \, h_0} \, .
\end{align}


\subsection{Prospects for intermediate operators ($h,\,\bar h\sim c$)}
\label{sec:inthhb}

We have seen that while the contribution of the extremely heavy 
operators in the chaos limit is suppressed, the contribution 
of the light operators alone is not enough to resolve the 
discrepancy between an individual block's growth rate and the 
chaos bound. Going beyond the light exchanged operators 
presents a challenge that we have not been able to resolve 
yet. We now offer some thoughts on how one might proceed with getting an 
estimate from the intermediate operators. 

As we have emphasized above, the resummation of the OPE requires three
pieces of data: the density of primaries, the OPE coefficients, and
the Virasoro blocks. In fact, the first two components can be
constrained by the results of \cite{Chang:2015qfa,Chang:2016ftb};
there it was shown that, under the assumption that the vacuum block
dominates throughout the Euclidean regime $0<z<\frac{1}{2}$ (which
follows at large $c$ from the bound used in \S\ref{sec:smallhhb}),
the OPE coefficients are fixed at leading order in large $c$ for all
$h > {\hat m}_v\, c$, where ${\hat m}_v\, c \sim \mathcal{O}(h_v)$ (cf., proposition $3$
in \cite{Chang:2015qfa}). While the precise expressions are slightly
involved, in principle we have a handle on the OPE coefficients
weighted by the density of states.

The primary technical obstacle is determining the semiclassical
Virasoro blocks for all exchanged operator dimensions, or at least the
behavior in the chaos region $q \to i$. These blocks are known to leading order for 
$\frac{h}{c} \ll 1$ (and $\frac{h_v}{c} \ll 1$), which we have significantly exploited above.
However,  as far as we are aware, there are no results available thus far for intermediate-to-heavy 
exchanged operators for times of order the scrambling time $t_*$. 

In recent years there has been some very interesting progress in understanding the recursive
determination of the pillow block $H_h(q)$, which allows for some
progress in this front. In particular, \cite{Kusuki:2017upd,
  Kusuki:2018wcv, Kusuki:2018nms, Kusuki:2018wpa} has found, at least
numerically and in some cases analytically, that the series coefficients in the $q$ expansion of $H_h(q)$,
\begin{align}
H_h(q) = \sum_{n=0}^\infty \mathfrak{h}_n \,q^n\, ,	
\end{align}
exhibit universal behavior in the $n\to \infty$ limit. For
example,\footnote{ For much more detailed investigation of the Virasoro
  blocks utilizing this and other similar asymptotic series data, we
  encourage the reader to consult \cite{Kusuki:2017upd,
    Kusuki:2018wcv, Kusuki:2018nms, Kusuki:2018wpa}.} if
$h_w \gg  \frac{3}{4}  \ca \gg h_v$, one has
\begin{equation}
\begin{split}
	\mathfrak{h}_{2k} \sim{}& ({-}1)^k (2k)^{\sf a} \;  e^{{\sf A} \, \sqrt{2k}} \quad 
	\text{as $k \to \infty$, with}  \\
	{\sf A} ={}& \pi \sqrt{\ca-2\,h_v}\, ,  \\
	{\sf a}  ={}& 2(h_v+h_w)- \frac{c+5}{8}\, .
\end{split}
\end{equation}
Here we recall that $\mathfrak{h}_n=0$ for odd $n$ in this OPE channel 
and $\ca$ is defined in \eqref{eq:ccdef}.

Using this asymptotic series data,  one can approximate the behavior of the block $H_h(q)$ near the boundary of the unit disk in the $q$ plane by treating the series summation as an integral to find
\begin{align}
\begin{split}
H_h(q) \sim{}& \int_0^\infty d k\, (2k)^{\sf a}  \; \tilde q^{2k} \; e^{{\sf A} \sqrt{2k}}  \\
	={}&  e^{{\sf A}^2/4\gamma} \; \gamma^{-{\sf a}  - \frac{3}{2}} 
	\bigg\{ \gamma^{\frac{1}{2}} \, \Gamma(1+{\sf a} ) \; {}_1 F_1\left({-}{\sf a} -
	\tfrac{1}{2}, \tfrac{1}{2}, {-} \frac{{\sf A}^2}{4\gamma}\right)   \\
	& \qquad \qquad  \qquad  \quad + {\sf A} 
	\, \Gamma(1+{\sf a} ) {}_1 F_1\left({-}{\sf a} , \tfrac{3}{2}, {-} 
	\frac{{\sf A}^2}{4\gamma}\right) \bigg\}  \\
	\sim{}& \frac{{\sf A}^{2{\sf a} +1}}{\gamma^{2{\sf a} +\frac{3}{2}}} \; 
	e^{{\sf A}^2/4\gamma} \, ,
\end{split}
\end{align}
where we have set $q = i \,\tilde q$, $\gamma = {-} \log 
\tilde q$, and taken the $c\to \infty$ limit in the last 
line. Sending $q\to i$, one can check that $\gamma \sim 
\frac{{-} \pi^2}{4\log (z/16)}$ to finally arrive at
\begin{align}
	H_h(q \to i) \sim \frac{{\sf A}^{2{\sf a} +1}}{\gamma^{2{\sf a} +\frac{3}{2}}}\, z^{{-}(\ca-2h_v)}\, .	
\end{align}
This power law behavior in $z$ essentially cancels the power law prefactor in the full Virasoro block ${\cal F}_h(z)$ (up to logarithmic corrections in $z$) and one is left with
\begin{align}
	\frac{{\cal F}_h(z)}{{\cal F}_0(z)} \to q^h\, .
\end{align}
From this, we see that the (holomorphic) contribution of the 
blocks to the four point function is exponentially 
suppressed, $z^{2h_v} {\cal F}_h \propto e^{{-}\frac{2\pi}
{\beta} 2h_vt}$, for $q$ asymptotically close to $q=i$.

Unfortunately, this asymptotic behavior is probing $q$ very 
close to the unit circle, which corresponds to very late 
times, while the chaotic behavior we are interested in is 
situated near the scrambling time. A simple way to see that 
we aren't quite seeing the physics we are interested in is 
that this limiting behavior is independent of the exchanged 
operator dimension $h$, whereas we expect there to be a 
delicate interplay among blocks near the scrambling time to 
be compatible with the chaos bound. It would be very 
interesting if one can push the asymptotic series 
coefficients further away from the unit circle to access 
times near the scrambling time $t_*$, but we leave this to 
future work.

\section{The AdS gravity story}
\label{sec:adsgrav}

The main thrust of our discussion has been to demonstrate that one needs to control contributions from operators with finite twist and large spin, to validate the truncation of the Euclidean correlator to the vacuum block, prior to analytically continuing to the chaos region. We now turn to analyzing the blocks directly from the bulk AdS dual, by examining the semiclassical gravitational picture. We briefly review the shockwave computation originally used to derive the chaos correlator in \cite{Shenker:2013pqa}, and show how to use a probe spinning particle to estimate the contribution from non-vacuum blocks.

Recall that when analytically continuing the Euclidean four-point function \eqref{eqn:EuclideanWWVV} to the OTO four-point function \eqref{OtO4-pt}, the imaginary time $\epsilon_i$'s in \eqref{eqn:imaginary_time} are assumed to be infinitesimal. The OTO four-point function is closely related to an expectation value in a thermofield double state \cite{Shenker:2013pqa},
\ie\label{eqn:vevInTfd}
\bra\psi V_L(0,x) V_R(0,x)\ket\psi
\fe
where the state $\psi$ is given by
\ie
\ket\psi=W_R(t+i\tau,x)\ket{\text{TFD}},~~~
\ket{\text{TFD}} = {1\over \sqrt{Z}}\sum_n e^{-{\beta\over 2}E_n }\ket{E_n}_L\otimes\ket{E_n}_R\,.
\fe
This observable is obtained by analytically continuing the Euclidean four-point function \eqref{eqn:EuclideanWWVV} with finite imaginary time $\epsilon_i$'s as\footnote{ More explicitly, we have
\ie
\bra\psi V_L V_R\ket\psi&= {1\over Z}\sum_{m,n}e^{-{\beta\over 2}(E_m+E_n) } \bra{E_m}_L\otimes\bra{E_m}_R W_R(t-i\tau) V_L V_R W_R(t+i\tau)\ket{E_n}_L\otimes\ket{E_n}_R
\\
&=\vev{W(t-i\tau)VW(t+i\tau)V(i\tfrac{\beta}{ 2})}_\beta.
\fe}
\ie
\epsilon_1=-\tau,\quad\epsilon_2=\tau,\quad\epsilon_3=0,\quad\epsilon_4={\B\over 2}.
\fe
$\tau$ is a regulating parameter replacing $\epsilon_{1,2}$ which will show up in the final result. 

The thermofield double state is holographically dual to the eternal AdS black hole (BTZ black hole) background, with two asymptotic boundaries that correspond to two copies of the CFT \cite{Maldacena:2001kr}. In the $h_w,c\gg h_v\gg1$ limit, the operators $W$'s create a shockwave in the BTZ black hole background.  The $V$ operators create a massive particle propagating in such background. The vacuum block contribution to the expectation value \eqref{eqn:vevInTfd} can be computed by evaluating the on-shell worldline action of the massive particle \cite{Roberts:2014ifa}. 

In general, the massive particle interacts with the shockwave beyond the minimal gravitational coupling. The non-minimal interactions originate from the exchanges of virtual particles, which generically carry nonzero masses and spins. The bulk exchange processes of virtual particles capture the contributions from the non-vacuum blocks to the expectation value \eqref{eqn:vevInTfd}. We verify this expectation by explicitly matching the non-vacuum blocks $G_0(x,t)G_{h, \bar h}(x,t)$ in \eqref{eqn:G0Gh} with the on-shell worldline actions for the massive particle created by the $V$ operators and the exchanged virtual particles. 

First, in the  \S\ref{sec:scalar_particle}, we consider the special case of scalar virtual particles and the non-vacuum blocks with $h=\bar h$. Subsequently, in  \S\ref{sec:spinning}, we introduce the worldline action for spinning particles, and match their on-shell action with spinning Virasoro blocks. Some of the details relevant for the computation can be found in the appendices. The BTZ shockwave metric is reviewed in Appendix \ref{sec:BTZshockwave}. The geodesic equations in the BTZ shockwave background are solved in Appendix \ref{sec:geodesics}.
We also use this technology to compute the semiclassical blocks in Euclidean signature in Appendix \ref{sec:semiwline} for completeness. 

\subsection{Scalar particle}
\label{sec:scalar_particle}

As reviewed in Appendix \ref{sec:BTZshockwave}, the BTZ shockwave metric can be constructed by gluing two halves of the BTZ metric in the Kruskal-Szekeres coordinate with the gluing condition \eqref{eqn:vShift}. We label a point on the left (right) boundary by $\vec x_{L(R)}=(t,x)_{L(R)}$ and a bulk point by $\vec{\bf x}=(u,v,x)$. The metric depends on one function $f(x)$ that depends on the energy injected by the shockwave into the spacetime. Following \cite{Roberts:2014ifa} we will parameterize this function as\footnote{
	As described in Appendix A of \cite{Roberts:2014ifa} $P$ is determined by computing the stress-energy carried by the shockwave. In writing this expression we have used the Brown-Henneaux relation $\frac{\ell_{AdS}}{G_N} = \frac{3}{2}\, c $ to facilitate comparison with the CFT computations above.} 
\begin{align}
f(x) = 1+ \frac{3\pi}{c}\,  P\, e^{-x} \,, \qquad P =\frac{2\,h_w}{\sin \tau}\, e^{t}\,.
\label{eqn:fxP}
\end{align}

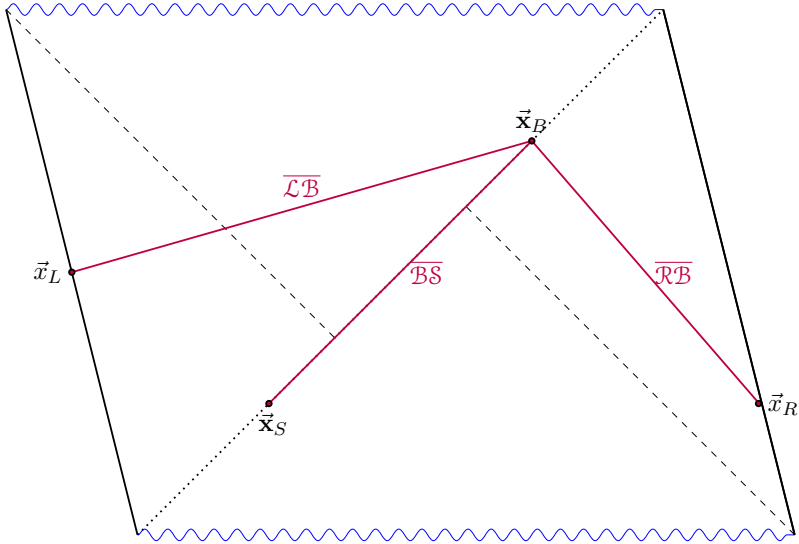
\begin{figure}[h]
\centering
\resizebox{0.7\textwidth}{!}{
\begin{tikzpicture}
\draw [dotted, thick, black] (-4,-4) -- (4,4);
\path [draw=blue,snake it]  (-4,-4) -- (6,-4) ;
\path [draw=blue,snake it]  (-6,4) -- (4,4) ;
\draw [thick, black] (-6,4) -- (-4,-4);
\draw [thick, black] (6,-4) -- (4,4);
\draw [thin, black, dashed] (6,-4) -- (1,1);
\draw [thin, black, dashed] (-6,4) -- (-1,-1);
\draw [thick, black] (6,-4) -- (4,4);
\draw [thick, purple]  (-5,0)  -- (-1.5,1) node[above]{$\overline{{\mathscr L} {\mathscr B}}$} -- (2, 2) -- (3.73,0) node[right]{$\overline{{\mathscr R} {\mathscr B}}$}  -- ( 5.46,-2);
\draw [thick, purple] (2, 2) -- (0,0) node[right]{$\overline{{\mathscr B} {\mathscr S}}$} --  (-2, -2) ;
\draw [thick, fill=purple] (-5,0) circle (0.25ex) node[left]{$\vec{x}_L$};
\draw [thick, fill=purple] (2,2) circle (0.25ex) node[above] {$\vec{\bf{x}}_B$ };
\draw [thick, fill=purple] (5.45,-2) circle (0.25ex) node[right] {$\vec{x}_R$}; 
\draw [thick, fill=purple] (-2,-2) circle (0.25ex) node[below]{$\ \vec{\bf x}_S$} ;
\end{tikzpicture}
}
\caption{ A sketch of the massive particle worldlines used for computing the chaos correlator from the bulk AdS geometry, depicted on the shockwave Penrose diagram.
}
\label{fig:shockPD}
\end{figure}
Let ${\mathscr L}$ be a point on the left boundary at $\vec x_L=(0,x)_L$, ${\mathscr R}$ be a point on the right boundary at $\vec x_R=(0,x)_R$, and $\mathscr B$ be a bulk point at $\vec{\bf x}_B=(0^-,v_B,x_B)\cong(0^+,v_B+f(x_B),x_B)$. The correlator we want is approximated by the on-shell action of a massive particle. The trajectory of the particle comprises of three segments:
propagation from the left boundary to the bulk $\overline{{\mathscr L} {\mathscr B}}$, 
propagation from a bulk point to the right boundary $\overline{{\mathscr R} {\mathscr B}}$, and a segment $\overline{{\mathscr B} {\mathscr S}}$ from ${\mathscr B}$ to a point ${\mathscr S}$ on the shockwave $\vec{\bf x}_S=(0^-,v_S,0)\cong(0^+,v_S+f(x_B),0)$. This is depicted in Fig.~\ref{fig:shockPD}. The action gets contribution from the three segments which are additive, viz., 
\ie\label{eqn:worldline}
&S=2\,h_v\, L(\vec x_L,\vec{\bf x}_B)+2\,h_v \, L(\vec x_R,\vec{\bf x}_B)+2\,h\, L(\vec {\bf x}_B,\vec{\bf x}_S),
\fe
where the function $L(\vec {\bf x},\vec{\bf y})$ is the geodesic distance between the points $\vec {\bf x}$ and $\vec{\bf y}$. The weight factors of $h_v$ and $h$ are assigned to the corresponding segments.   

The proper lengths of the worldlines $\overline{{\mathscr L} {\mathscr B}}$, $\overline{{\mathscr R} {\mathscr B}}$, and $\overline{{\mathscr B} {\mathscr S}}$ are explicitly given by
\ie
&L(\vec x_L,\vec{\bf x}_B)=\log\left[\cosh (x_B-x) - v_B\right],
\\
&L(\vec x_R,\vec{\bf x}_B)=\log\left[\cosh (x_B-x)+v_B+f(x_B)\right],
\\
&L(\vec {\bf x}_B,\vec{\bf x}_S)=|x_B|.
\fe
Without loss of generality, by assuming $x>0$, and the worldline action is extremized at
\ie\label{eqn:minimizedValues}
v_B=-{1\over 2}f(x_B),~~~e^{x_B}&=\sqrt{{2h_v-h\over 2h_v+h}\; e^{2x} \left(1+ f(x) \right) } \,.
\fe
We then have after plugging in the expressions for $f(x)$ from \eqref{eqn:fxP} 
\ie\label{eqn:minimizedAction}
e^{-S}&= \frac{\left(2\,h_v+h\right)^{h+2\,h_v} }{\left(2\,h_v-h\right)^{h-2\,h_v}\, \left( 4\, h_v^2\right)^{2\,h_v}}
\frac{e^{-2\, h \, x}}{ \left(  1+ \frac{6\pi \, h_w}{c\, \sin \tau} \,  e^{t-x}\right)^{h+2h_v}} \,,
\fe
which is consistent with CFT result, viz.,  
\ie
e^{-S}\propto G_0(x,t)G_{h, h}(x,t),
\fe
where $G_0(x,t)$ and $G_{h, \bar h}(x,t)$ are in \eqref{eqn:G0Gh}. This is the desired generalization of the result in \cite{Roberts:2014ifa} for $h\neq0$.

In our configuration, the worldline action \eqref{eqn:worldline} is independent of $v_S$. To determine $v_S$, we move the point $\vec {\bf x}_S$ infinitesimally along the $u$-direction. The resulting worldline action is minimized at
\ie
v_S=v_B\cosh x_B.
\fe

\subsection{Spinning particle}
\label{sec:spinning}

The motion of a spinning particle moving curved spacetime is described by the Mathisson-Papapetrou-Dixon equations \cite{Mathisson:1937zz,Papapetrou:1951pa,Dixon:1970zz}. In three dimensions, the Mathisson-Papapetrou-Dixon equation can be derived from minimizing the action \cite{Castro:2014tta}
\ie\label{eqn:SPA}
S=&\int ds\left({\mathfrak m}\sqrt{g_{\m\n}\,\dot x^\m \dot x^\n}+{\mathfrak s}\,\tilde n\cdot\nabla n\right)
\\
&+\int ds\left[\lambda_1 n\cdot\tilde n+\lambda_2n\cdot\dot{\vec x}+\lambda_3\tilde n\cdot\dot {\vec x}+\lambda_4(n^2+1)+\lambda_5(\tilde n^2-1)\right],
\fe
where $\nabla$ is the pullback of the covariant derivative, i.e. $\nabla=\dot x^\m\nabla_\m$. The $\lambda_i$'s on the second line are Lagrange multipliers that impose the constraints,
\ie\label{eqn:SAC}
n\cdot\tilde n=n\cdot\dot {\vec x}=\tilde n\cdot\dot {\vec x}=0,~~~n^2=-\tilde n^2=-1,
\fe
which imply that the spinning particle is moving in the space directions i.e. $|\dot {\vec x}|^2=1$.\footnote{ The action for a spinning particle moving in the time direction is given by modifying the $n^2+1$ on the second line of \eqref{eqn:SPA} to $n^2-1$.} Varying the action with respect to $x^\m$ gives the Mathisson-Papapetrou-Dixon equation
\ie\label{eqn:MPDeq}
\nabla({\mathfrak m}\,\dot x^\m + \dot x_\rho \nabla s^{\m\rho})+{1\over 2}\dot x^\n\, s^{\rho\sigma}\,R^\m{}_{\n\rho\sigma}=0,
\fe
where the spin tensor $s^{\m\n}$ is given by
\ie
s^{\m\n}={\mathfrak s}\,(n^\m\tilde n^\n-n^\n\tilde n^\m).
\fe
On the other hand, varying the action with respect to $n$ and $\tilde n$ does not lead to dynamical equations, as shown in \cite{Castro:2014tta}. In fact, the action \eqref{eqn:SPA} is insensitive to small variations of the normal vectors $n$ and $\tilde n$ along the trajectory, up to boundary terms \cite{Castro:2014tta}.

On a locally AdS$_3$ metric, the constraints \eqref{eqn:SAC} give 
\ie\label{eqn:stnsolved}
s_{\m\n}&={\mathfrak s}\,\epsilon_{\m\n\rho}\, \dot{x}^\rho,\quad n_\m&=-\epsilon_{\m\n\rho}\,\dot x^\n \,\tilde n^\rho,
\fe
where $\epsilon_{\m\n\rho}$ is a totally antisymmetric tensor with $\epsilon_{012}=\sqrt{-g}$. The equation of motion \eqref{eqn:MPDeq} becomes
\ie\label{eqn:AdS3EOM}
\nabla({\mathfrak m}\,\dot x^\m -{\mathfrak s}\,\epsilon^{\m\n\rho}\, \dot x_\n \nabla \dot x_\rho)=0,
\fe
which, in particular, admits geodesics, that satisfy $\nabla \dot x^\m=0$, as solutions.

The geodesic equations on the BTZ shockwave background are solved in Appendix \ref{sec:geodesics}. We evaluate the action \eqref{eqn:SPA} on the geodesic $\overline{{\mathscr B}{\mathscr S}}$. The first term in \eqref{eqn:SPA} gives the same result as in \S\ref{sec:scalar_particle} for scalar particles. The terms on the second line of \eqref{eqn:SPA} all vanish when the constraints \eqref{eqn:SAC} are satisfied.  Using \eqref{eqn:stnsolved}, the second term in \eqref{eqn:SPA} becomes
\ie\label{eqn:SPAspinterm}
S_{spin}=\int ds\,{\mathfrak s} \,\epsilon_{\m\n\rho}\, \dot x^\m \, \tilde n^\n\,\nabla \tilde n^\rho.
\fe
Consider covariantly constant normal vectors $q^\m(s)$ and $\tilde q^\m(s)$, which satisfy
\ie\label{eqn:qtqconditions}
&\nabla q=\nabla\tilde q=0,
\\
&q^2=-\tilde q^2=-1,~~~q\cdot\dot {\vec x}=\tilde q\cdot\dot {\vec x}=q\cdot\tilde q=0.
\fe
We can expand $\tilde n^\m(s)$ in terms of $q^\m(s)$ and $\tilde q^\m(s)$ as
\ie
\tilde n(s)=\cosh(\eta(s))\,q(s)+\sinh(\eta(s))\, \tilde q(s).
\fe
The integral \eqref{eqn:SPAspinterm} becomes
\ie\label{eqn:SspinEta}
S_{spin}=\int ds\,{\mathfrak s}\, \dot\eta(s)={\mathfrak s}\,(\eta(s_f)-\eta(s_i)),
\fe
which can be also written as
\ie
S_{spin}={\mathfrak s}\log\left(q(s_f)\cdot \tilde n_f-\tilde q(s_f)\cdot \tilde n_f\over q(s_i)\cdot \tilde n_i-\tilde q(s_i)\cdot \tilde n_i\right),
\fe
where $\tilde n_i=\tilde n(s_i)$ and $\tilde n_f=\tilde n(s_f)$ are the initial and final values of $\tilde n(s)$.
The trajectory of this geodesic $\overline{{\mathscr B} {\mathscr S}}$ and the covariantly constant normal vectors $q(s)$ and $\tilde q(s)$ are given explicitly in \eqref{eqn:trajectorySB}, \eqref{eqn:normalQ} and \eqref{eqn:normalTq}. Let us define $\tilde n_i=(\tilde n_i^u,\tilde n_i^v,\tilde n_i^x)$ and $\tilde n_f=(\tilde n_f^u,\tilde n_f^v,\tilde n_f^x)$, and we have
\ie
S_{spin}={\mathfrak s}\log\left(\tilde n_f^u\over \tilde n_i^u\right).
\fe
The orthonormal conditions \eqref{eqn:SAC} on the normal vectors $\tilde n_i$ and $\tilde n_f$ give
\ie\label{eqn:orthNiNf}
&\tilde n_i^x=0,\quad -4\tilde n^u_{i}\tilde n^v_{i}=1,\quad \tilde n_f^x=2\tilde n_f^uv_B\sinh|x_B|,\quad -4\tilde n^u_f\tilde n^v_f+(\tilde n^x_f)^2=1.
\fe

The above equations do not uniquely determine $\tilde n_i$ and $\tilde n_f$. One need to specify boundary conditions on the normal vectors $n$ and $\tilde n$. A natural boundary condition at a cubic vertex is such that the normal vector $n$ is perpendicular to the two-plane spanned by the three velocity vectors at the cubic vertex. By the conditions $n\cdot\tilde n=0$, the normal vector $\tilde n_i$ is inside the two-plane. 
By this boundary condition, the normal vector $\tilde n_i$ is proportional to the difference of the velocity vectors \eqref{eqn:leftVelocityAtB} and \eqref{eqn:rightVelocityAtB} of the geodesics $\overline{{\mathscr R} {\mathscr B}}$ and $\overline{{\mathscr L} {\mathscr B}}$ at the point ${\mathscr B}$. Explicitly, we have
\ie\label{eqn:aveVelocity}
&\tilde n_i^u={1\over 2\sqrt{1+f(x)}},
\\
&\tilde n_i^v=-{1\over 2}\sqrt{1+f(x)},
\\
&\tilde n_i^x=0.
\fe
We do not have a general prescription for the boundary condition for a spinning particle trajectory ending on a shockwave. We will simply pick the $\tilde n_f$ to be a constant (independent of $x$). The action $S_{spin}$ is
\ie
S_{spin}={1\over 2}(h-\bar h)\log\left(1+f(x)\right)+{\rm const.}
\fe
Again, we find
\ie
e^{-S}\propto G_0(x,t)G_{h, \bar h}(x,t),
\fe
where the Virasoro block \eqref{eqn:G0Gh} with ${\mathfrak s}=h-\bar h\neq 0$ can be written as
\ie
G_0(x,t)G_{h,\bar h}(x,t)\propto \left(\frac{6\pi \, h_w}{c\, \sin\tau}\, e^t\right)^{h-\bar h}
\frac{e^{- (h+\bar h) \, |x| }}{
	 \left(  1+ \frac{6\pi \, h_w}{c\, \sin \tau} \,  e^{t-|x|}\right)^{2\, h_v + h}
}
\fe

\section{Discussion}
\label{sec:discuss}

Motivated by the need to characterize explicitly the constraints on large central charge CFTs to admit a classical gravitational dual, we examined the OTO 4-point function which probes ergodic dynamics of the theory. We argued while the Lyapunov exponent attains its maximal value when one focuses on the contribution from the vacuum block alone, this does not suffice, as individual non-vacuum blocks make comparable or even larger contribution around the timescale of interest. This suggests an intricate conspiracy between non-vacuum blocks, which is eminently possible. We sought to understand whether this conspiracy could be quantified in terms of bounds on the OPE data, but lack of control on the generic OPE coefficients and the blocks themselves, prevented us from making quantitative statements. The results presented here should be viewed as a first step in a program of corralling the non-vacuum block contributions in the Regge regime in two dimensional CFTs.

\begin{figure}[h]
\centering
\resizebox{0.7\textwidth}{!}{
\begin{tikzpicture}
\draw [thick, black, ->] (0,0) -- (10,0) node[right]{$h$};
\draw [thick, black, ->] (0,0) -- (0,10) node[above]{$\bar{h}$};
\draw[thick, black]  (2,0) arc (0:90:2);
\draw[thick,dotted] (4,0) arc (0:90:4); 
\draw[thick, black] (3.65,3.65) node[below,rotate=-45]{intermediate operators};
\draw[thick,black,dashed] (6,0) arc (0:35:6);
\draw[thick,black,dashed] (0,6) arc (90:55:6);
\draw[thick, black,dashed] (4.91,3.44) -- (10,3.44);
\draw[thick, black,dashed] (3.44,4.91) -- (3.44,10);
\draw[thick, black,dotted] (4.25,4.25) -- (5,10); 
\draw[thick, black,dotted] (4.25,4.25) -- (10,5); 
\draw[thick, black] (2,0) node[below]{$h_{\text{gap}}$};
\draw[thick, black] (0,2) node[left]{$\bar{h}_{\text{gap}}$};
\draw[thick, black] (0,6) node[left]{$\bar{h}_{0} + \ca$};
\draw[thick, black] (6,0) node[below]{$h_{0} +\ca$};
\draw[thick, black] (4,0) node[below]{$\epsilon\, c$};
\draw[thick, black] (0,4) node[left]{$\epsilon\,c$};
\node[label={[label distance=0.5cm,text depth=-1ex,rotate=45]right: heavy operators}] at (4.75,6) {};
\node[label={[label distance=0.5cm,text depth=-1ex,rotate=45]right: $\sim e^{-\frac{2\pi}{\beta}(\bar{h}_0 + \ca - 2\Delta_v)\,t}$}] at (5.5,5.5) {};
\node[label={[label distance=0.5cm,text depth=-1ex,rotate=0]right: large spin}] at (6,2.25) {};
\node[label={[label distance=0.5cm,text depth=-1ex,rotate=0]right: $\sim e^{{-}\frac{2\pi}{\beta}\bar h_m t -\frac{ \pi\beta}{8\,t} \, h_0}$}] at (6,1.25) {};
\node[label={[label distance=0.5cm,text depth=-1ex,rotate=90]right: large spin }] at (1.25,6) {};
\node[label={[label distance=0.5cm,text depth=-1ex,rotate=90]right: $\sim e^{\frac{2 \pi}{\beta}(h_m t_* - \bar h_0 t)}$ }] at (2.25,6) {};
\draw [thick, black] (-0.1,0.5) node[right]{$\sim e^{\frac{2\pi}{\beta} (t-t_*)}$};
\draw [thick, black] (1.4,2)  node[right]{$\sim e^{\ell_\text{gap} t}$};
\end{tikzpicture}
}
\caption{ A rough characterization of the results for the chaos correlator (temporal dependence alone) in various domains analyzed in the paper.
}
\label{fig:domains}
\end{figure}
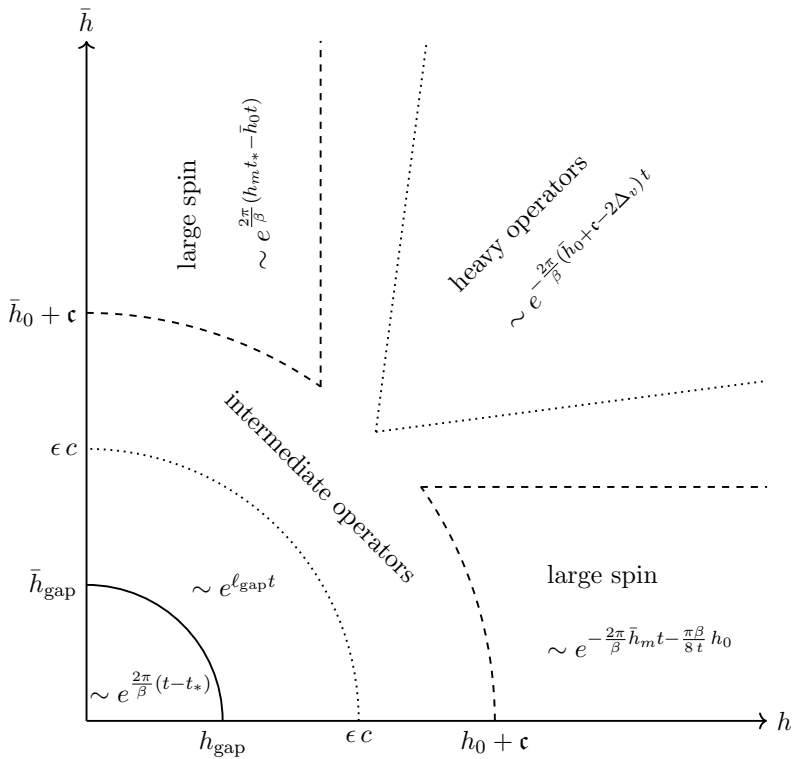

There are several approaches one could take to better the analysis presented herein. In higher dimensions, the Lorentzian inversion formula obtained in \cite{Caron-Huot:2017vep} provides us with a useful way to view the contributions from higher twist operators to the global conformal blocks. For Virasoro blocks one is stymied by the presence of twist-zero operators in carrying out the inversion. One can however attempt to use the analogous gadget for 2d CFTs, the Virasoro fusion kernel, as recently explored in \cite{Collier:2018exn}. We have undertaken some preliminary explorations in this front in \S\ref{sec:estimates}, especially to bound the contribution of the very heavy and large spin operators, but the interesting regime where there is a non-trivial interplay between light and intermediate operators remains to be better understood. Fig.~\ref{fig:domains} displays a summary of the various regimes where we have estimates for the contributions to the chaos correlator.

It would also be useful to examine the OTO correlator in non-rational CFTs explicitly (even for moderate central charge) to understand better the contribution  from the non-vacuum blocks. One could for example dissect the SYK family of models explored in \cite{Murugan:2017eto,Bulycheva:2018qcp,Peng:2018zap}  to understand better how the non-vacuum blocks conspire to bring down the chaos exponent to a sub-maximal value. For instance, the models explored in \cite{Peng:2018zap}  admit a one-parameter extension where one can tune the Lyapunov exponent between zero (the integrable limit) and a sub-maximal value. One can imagine being able to discern how higher spin states start getting lifted as we tune away from the integrable point. These exercises would be interesting to carry out and hopefully will teach us how we can assemble a list of criteria for putting together a holographic CFT.

\acknowledgments

It is a pleasure to thank  Jared Kaplan, Henry Maxfield, and Douglas Stanford for useful discussions. We would like to thank KITP, UCSB for hospitality during the workshop ``Chaos and Order: From strongly correlated systems to black holes'', where the research was supported in part by the National Science Foundation under Grant No. NSF PHY17-48958 to the KITP.
The authors are supported  by U.S.\ Department of Energy grant DE-SC0009999 and by funds from the University of California. 

\newpage
\appendix

\section{BTZ shockwave metric}\label{sec:BTZshockwave}

The non-rotating BTZ metric in the Kruskal-Szekeres coordinate is given by (setting $\ell_{AdS} =1$)
\ie\label{eqn:BTZKS}
ds^2=-{4\over (1+uv)^2} du dv+ r_+^2\, {(1-uv)^2\over (1+uv)^2}dx^2.
\fe
In this coordinate, the horizon is at $uv=0$, and the future and past singularities are at $uv=1$. There are two disjoint boundaries at $uv=-1$. Under the coordinate transformations
\ie\label{eqn:uvToRt}
u=\mp {r-r_+\over \sqrt{r^2-r_+^2}}e^{r_+\,t},~~~v=\pm {r-r_+\over \sqrt{r^2-r_+^2}}e^{-r_+\,t},
\fe
the asymptotic regions $v\ge0\ge u$ and $v\le 0\le u$ can be transformed to the more standard BTZ blackhole metric,
\ie
ds^2=-(r^2-r_+^2)dt^2+{1\over (r^2-r_+^2)}dr^2+r^2dx^2.
\fe
For computational convenience, we will sometimes use the Poincar\'e coordinate,
\ie\label{eqn:Poincare_patch}
ds^2={dy^2+dzd\bar z\over y^2},
\fe
which is related to the BTZ metric by
\ie\label{eqn:wzToRt}
z={\sqrt{r^2-r_+^2}\over r}e^{r_+\,(x+t)},~~~\bar z={\sqrt{r^2-r_+^2}\over r}e^{r_+\,(x-t)},~~~~y={r_+\over r}e^{r_+\,x}.
\fe

Next, let us introduce a shockwave, which is a delta function source of energy momentum tensor localized at $u=x=0$,
\ie
T_{uu}(u,v,x)=P\, \delta(u)\delta(x).
\fe
The metric ansatz of the shockwave is given by
\ie\label{eqn:shockwaveMetric}
ds^2=-{4\over (1+uv)^2} du dv+{(1-uv)^2\over (1+uv)^2}dx^2+4 \, \delta(u)\, f(x) \, du^2,
\fe
where we have assumed $r_+=1$ for simplicity. The Einstein's equations $G_{\m\n}-g_{\m\n}=8\pi G_N T_{\m\n}$ gives (using $G_N = \frac{3}{2\,c}$)
\ie
f(x)= \frac{3\pi}{c}  P\, e^{-|x|}.
\fe
Equivalently, the shockwave geometry can also be realized by gluing the $u<0$ and $u>0$ parts of BTZ metric with a glueing condition \cite{Shenker:2013pqa,Roberts:2014isa}
\ie\label{eqn:vShift}
v\big|_{u<0}\cong v\big|_{u>0}+f(x).
\fe

\section{Geodesic equations}
\label{sec:geodesics}

In this section, we solve the geodesic equations in the shockwave background \eqref{eqn:shockwaveMetric}.

\subsection{Away from the shockwave}

Away from the shockwave $u=0$, the geometry is locally AdS$_3$ and can be described by the Poincar\'e metric \eqref{eqn:Poincare_patch}. We will work in the region $v\ge0\ge u$. The results for the region $v\le0\le u$ can be obtained by first flipping the signs of both the $u$ and $v$, and shifting the $v$-coordinate by \eqref{eqn:vShift}.

The geodesic equation in the Poincar\'e patch of AdS$_3$ is
\ie
&\ddot z-{2\over y}\dot y\dot z=0,
\\
&\ddot {\bar z}-{2\over y}\dot y\dot {\bar z}=0,
\\
&\ddot y+{1\over y}(\dot z\dot{\bar z}-\dot y^2)=0,
\fe
where the dot `$\cdot$' denotes derivative with respect to the proper distance $s$ normalized by the equation
\ie\label{eqn:NorS}
\dot z\dot{\bar z}+\dot y^2=y^2.
\fe
The solution to the geodesic equation is given by
\ie\label{eqn:geodesic_in_Poincare}
&y(s)={y_iy_f\sinh(s_f-s_i)\over y_i\sinh(s-s_i)-y_f\sinh(s-s_f)},
\\
&z(s)=z_i+(z_f-z_i){y_i\sinh(s-s_i)\over y_i\sinh(s-s_i)-y_f\sinh(s-s_f)},
\\
&\bar z(s)=\bar z_i+(\bar z_f-\bar z_i){y_i\sinh(s-s_i)\over y_i\sinh(s-s_i)-y_f\sinh(s-s_f)}.
\fe

Integrating the equation \eqref{eqn:NorS} gives the geodesic distance between the points $\vec{\bf x}_i=(z_i,\bar z_i,y_i)$ and $\vec{\bf x}_f=(z_f,\bar z_f,y_f)$,
\ie\label{eqn:geodesicDistance}
L(\vec{\bf x}_i,\vec{\bf x}_f)\equiv s_f-s_i=\cosh^{-1}\left[1+{(y_f-y_i)^2+(z_f-z_i)(\bar z_f-\bar z_i)\over 2 y_i y_f}\right].
\fe
Using the coordinate transformations \eqref{eqn:uvToRt} and \eqref{eqn:wzToRt}, we find the formula for the geodesic distance in the Kruskal-Szekeres coordinate,
\ie
L(\vec{\bf x}_i,\vec{\bf x}_f)=\cosh^{-1}\left[{2(u_i v_f+u_f v_i)+(1-u_i v_i)(1-u_f v_f)\cosh(x_f-x_i)\over (1+u_i v_i)(1+u_f v_f)}\right].
\fe
The geodesic distance diverges when we take the point $\vec{\bf x}_i$ approaching the boundary,
\ie
L(\vec{\bf x}_i,\vec{\bf x}_f)=\log (2r_i)+\log\left[{(1-u_f v_f)\cosh(x_f-x_i)+ u_f e^{-t_i}- v_f e^{t_i}\over 1+u_f v_f}\right] + {\cal O}(r_i^{-1}),
\fe
where the $r_i$ and $t_i$ are the coordinates  defined in \eqref{eqn:uvToRt}. We define the regularized geodesic distance of a boundary point $\vec x_i=(x_i, t_i)$ and a bulk point $\vec{\bf x}_f=(u_f,v_f,x_f)$ as
\ie\label{eqn:bdryBulkGeodesic}
L(\vec{ x}_i,\vec{\bf x}_f)=\log\left[{(1-u_f v_f)\cosh(x_f-x_i)+ u_f e^{-t_i}- v_f e^{t_i}\over 1+u_f v_f}\right].
\fe

Finally, the geodesic from a boundary point $\vec x_i=(z_i,\bar z_i)$ to a bulk point $\vec{\bf x}_f=(z_f,\bar z_f,y_f)$ can be described by
\ie\label{eqn:boundaryToBulk}
&y(s)=y_f{e^{s-s_f}(|z_i-z_f|^2+y_f^2)\over e^{2(s-s_f)}|z_i-z_f|^2+y_f^2},
\\
&z(s)=z_i\left(1-{e^{2(s-s_f)}(|z_i-z_f|^2+y_f^2)\over e^{2(s-s_f)}|z_i-z_f|^2+y_f^2}\right)+z_f{e^{2(s-s_f)}(|z_i-z_f|^2+y_f^2)\over e^{2(s-s_f)}|z_i-z_f|^2+y_f^2},
\\
&\bar z(s)=\bar z_i\left(1-{e^{2(s-s_f)}(|z_i-z_f|^2+y_f^2)\over e^{2(s-s_f)}|z_i-z_f|^2+y_f^2}\right)+\bar z_f{e^{2(s-s_f)}(|z_i-z_f|^2+y_f^2)\over e^{2(s-s_f)}|z_i-z_f|^2+y_f^2},
\fe
which is obtained by eliminating the $s_i$ dependence in \eqref{eqn:geodesic_in_Poincare} by using \eqref{eqn:geodesicDistance}, and taking the $y_i\to 0$ limit.

\subsection{Crossing the shockwave}

In the section, we study the geodesic equations for a geodesic that crosses the shockwave at $u=0$. The geodesic equations on the shockwave background \eqref{eqn:shockwaveMetric} are
\ie
&\ddot u-{2v\over 1+uv}\dot u^2-{u(1-uv)\over 1+uv}\dot x^2=0,
\\
&\ddot v-{2u\over 1+uv}\dot v^2-{v(1-uv)\over 1+uv}\dot x^2-\left[2v\delta(u)+\delta'(u)\right]f(x)\dot u^2-2\delta(u)f'(x)\dot x\dot u=0,
\\
&\ddot x-{4v\over 1+u^2v^2}\dot x\dot u-{4u\over 1+u^2v^2}\dot x\dot v-2\delta(u)f'(x)\dot u^2=0,
\fe
and the normalization of the proper distance gives
\ie\label{eqn:norDistanceShock}
-{4\over (1+uv)^2} \dot u\dot v+{(1-uv)^2\over (1+uv)^2}\dot x^2+4\delta(u)f(x)\dot u^2=1.
\fe

First, using \eqref{eqn:norDistanceShock}, we find that the $v$-coordinate jumps at the shockwave at $u=0$ by the amount of
\ie\label{eqn:jump_V}
\delta v\equiv\int_{-\epsilon}^{\epsilon}\dot v{du\over \dot u}=\int_{-\epsilon}^{\epsilon} \delta(u)f(x)du=f(x),
\fe
which agrees with the gluing condition \eqref{eqn:vShift}. When $f(x)$ is not a constant, the velocity vector $(\dot u,\dot v,\dot x)$ of a geodesic also jumps when crossing the shock wave,
\ie\label{eqn:jump_dU_dV_dX}
\delta \dot u&\equiv\int_{-\epsilon}^{\epsilon}\ddot u{du\over \dot u}=0,
\\
\delta \dot x&\equiv\int_{-\epsilon}^{\epsilon}\ddot x{du\over \dot u}=\int_{-\epsilon}^{\epsilon}2\delta(u)f'(x)\dot u du=2f'(x)\dot u\big|_{u=0},
\\
\delta \dot v&\equiv\int_{-\epsilon}^{\epsilon}\ddot v{du\over \dot u}=\int_{-\epsilon}^{\epsilon}\left\{\left[2v\delta(u)+\delta'(u)\right]f(x)\dot u+2\delta(u)f'(x)\dot x\right\} du
\\
&=2v_Bf(x)\dot u-f'(x)\left(\dot x+{1\over 2}\delta\dot x\right)-f(x){\ddot u\over \dot u}+2f'(x)\left(\dot x+{1\over 2}\delta\dot x\right)
\\&
=\left[f'(x)\dot x+f'(x)^2\dot u\right]\big|_{u=0}.
\fe
In the Appendix \ref{sec:geoLB_RB_BS}, we will consider a geodesic that crosses halfway through the shockwave. The amount of jump is given by \eqref{eqn:jump_V} and \eqref{eqn:jump_dU_dV_dX} with $f(x)$ and $f'(x)$ replaced by ${1\over 2}f(x)$ and ${1\over 2}f'(x)$,
\ie
\delta_{1\over 2} v&\equiv \delta v\big|_{f\to{1\over 2}f}={1\over 2}f(x),
\\
\delta_{1\over 2} \dot u&\equiv \delta \dot u\big|_{f'\to{1\over 2}f'}=0,
\\
\delta_{1\over 2} \dot v&\equiv \delta \dot v\big|_{f'\to{1\over 2}f'}=\left[{1\over 2}f'(x)\dot x+{1\over 4}f'(x)^2\dot u\right]\big|_{u=0},
\\
\delta_{1\over 2} \dot x&\equiv \delta \dot x\big|_{f'\to{1\over 2}f'}=f'(x)\dot u\big|_{u=0}.
\fe

\subsection{Three geodesic segments for the correlator}
\label{sec:geoLB_RB_BS}

\paragraph{1. $\overline{{\mathscr L}{\mathscr B}}$ geodesic} 
The trajectory of this geodesic is given by the \eqref{eqn:boundaryToBulk} with
\ie
&z_i=e^x,~~~\bar z_i=e^x, ~~~y_i=0,
\\
&z_f=0,~~~\bar z_f=2v_Be^{x_B},~~~y_f=e^{x_B}.
\fe
We are interested the velocity vector at the bulk point ${\mathscr B}$. We first compute the velocity vector of the geodesic at $s=s_f$,
\ie
&\dot u(s_f)={1\over 2(\cosh(x-x_B)-v_B)},
\\
&\dot v(s_f)=v_B-{1-v_B^2\over 2(\cosh(x-x_B)-v_B)},
\\
&\dot x(s_f)=-{\sinh(x-x_B)\over\cosh(x-x_B)- v_B}.
\fe
Next we need to take into account the jumps in the velocity vector at the shockwave. Because the point ${\mathscr B}$ is on the shockwave at $u=0$, we consider the geodesic that is just halfway crossing the shockwave. Therefore, the velocity vector at point ${\mathscr B}$ is given by
\ie\label{eqn:leftVelocityAtB}
&\dot u\big|_{{\mathscr B}}=\dot u(s_f)+\delta_{1\over 2} \dot u={1\over 2(\cosh(x-x_B)-v_B)},
\\
&\dot v\big|_{{\mathscr B}}=\dot v(s_f)+\delta_{1\over 2} \dot v=v_B-{1-v_B^2+f'(x_B)\sinh(x-x_B)-{1\over 4}f'(x_B)^2\over 2(\cosh(x-x_B)-v_B)},
\\
&\dot x\big|_{{\mathscr B}}=\dot x(s_f)+\delta_{1\over 2} \dot x={{1\over 2}f'(x_B)-\sinh(x-x_B)\over\cosh(x-x_B)- v_B},
\fe
where $\delta_{1\over 2} \dot u$, $\delta_{1\over 2} \dot v$, and $\delta_{1\over 2} \dot x$ are given in \eqref{eqn:jump_dU_dV_dX}.

\paragraph{2. $\overline{{\mathscr R}{\mathscr B}}$ geodesic}  The trajectory of this geodesic is simply given by flipping the signs of both $u$ and $v$ of the geodesic $\overline{{\mathscr L}{\mathscr B}}$. The velocity vector at $s=s_f$ is given by
\ie
&\dot u(s_f)=-{1\over 2(\cosh(x-x_B)+v_B+f(x_B))},
\\
&\dot v(s_f)=v_B+f(x_B)+{1-v_B^2\over 2(\cosh(x-x_B)+v_B+f(x_B))},
\\
&\dot x(s_f)=-{\sinh(x-x_B)\over \cosh(x-x_B)+v_B+f(x_B)}.
\fe
The velocity vector at point ${\mathscr B}$ is given by
\ie\label{eqn:rightVelocityAtB}
&\dot u\big|_{{\mathscr B}}=-{1\over 2(\cosh(x-x_B)+v_B+f(x_B))},
\\
&\dot v\big|_{{\mathscr B}}=v_B+f(x_B)+{1-(v_B+f(x_B))^2+f'(x_B)\sinh(x-x_B)-{1\over 4}f'(x_B)^2\over 2(\cosh(x-x_B)+v_B+f(x_B))},
\\
&\dot x\big|_{{\mathscr B}}={{1\over 2}f'(x_B)-\sinh(x-x_B)\over \cosh(x-x_B)+v_B+f(x_B)}.
\fe

\paragraph{3. $\overline{{\mathscr B}{\mathscr S}}$ geodesic}
The trajectory of this geodesic is given by the \eqref{eqn:boundaryToBulk} with
\ie
&z_i=0,~~~\bar z_i=2v_Be^{x_B},~~~y_i=e^{x_B},
\\
&z_f=0,~~~\bar z_f=2v_S,~~~y_f=1.
\fe
The geodesic distance \eqref{eqn:geodesicDistance} is
\ie
s_f-s_i=|x_B|.
\fe
We can conveniently choose $s_i=0$ and $s_f=|x_B|$. In the Kruskal-Szekeres coordinate, we have
\ie\label{eqn:trajectorySB}
&x(s)=x_B\pm s,\quad u(s)=0,
\\
&v(s)=\left[v_S\sinh s-v_B\sinh(s-|x_B|)\right]\,{\rm csch}\,|x_B|,
\fe
where the plus (minus) sign is for $x_B\le0$ $(x_B>0)$. The velocity vector at point ${\mathscr B}$ is
\ie\label{eqn:middleVelocityAtB}
 \dot u\big|_{{\mathscr B}}=\dot u(s_i)=0,\quad \dot v\big|_{{\mathscr B}}=\dot v(s_i)=0,\quad \dot x\big|_{{\mathscr B}}=\dot x(s_i)=\pm 1.
\fe
A consistent check of our solutions is that the momentum conservation is satisfied at point ${\mathscr B}$, i.e.
\ie
2h_v \dot{\vec x}_{\overline{{\mathscr L}{\mathscr B}}}\big|_{{\mathscr B}}+2h_v \dot{\vec x}_{\overline{{\mathscr R}{\mathscr B}}}\big|_{{\mathscr B}}=2h \dot{\vec x}_{\overline{{\mathscr B}{\mathscr S}}}\big|_{{\mathscr B}},
\fe
where $\dot{\vec x}_{\overline{{\mathscr L}{\mathscr B}}}\big|_{{\mathscr B}}$, $\dot{\vec x}_{\overline{{\mathscr R}{\mathscr B}}}\big|_{{\mathscr B}}$, $\dot{\vec x}_{\overline{{\mathscr R}{\mathscr S}}}\big|_{{\mathscr B}}$ are given in \eqref{eqn:leftVelocityAtB}, \eqref{eqn:rightVelocityAtB}, and \eqref{eqn:middleVelocityAtB}, and we have used \eqref{eqn:minimizedValues}.

Let us consider the normal vectors $q(s)$ and $\tilde q(s)$ that satisfy the orthonormal conditions
\ie
&q\cdot \tilde q=q\cdot \dot{\vec x}=\tilde q\cdot \dot{\vec x}=0,\quad-q^2=\tilde q^2=1,
\fe
and the parallel transport equations
\ie
\nabla q=\nabla \tilde q=0.
\fe
The equation $\nabla q=0$ can be written explicitly as
\ie
&\dot q^u=0,
\\
&\dot q^v-v \dot x q^x=0,
\\
&\dot q^x-2v \dot x q^u=0.
\fe
The solution to the orthonormal conditions and the parallel transport equations is
\ie\label{eqn:normalQ}
& q^x(s)=2 q^u \left[v_S\cosh s-v_B\cosh(s-|x_B|)\right]\,{\rm csch}\,x_B,
\\
& q^v(s)={1\over 4q^u} +q^u\left[v_S\cosh s-v_B\cosh(s-|x_B|)\right]^2\,{\rm csch}^2\,x_B,
\fe
and
\ie\label{eqn:normalTq}
&\tilde q^x(s)=2 \tilde q^u \left[v_S\cosh s-v_B\cosh(s-|x_B|)\right]\,{\rm csch}\,x_B,
\\
&\tilde q^v(s)=-{1\over 4\tilde q^u} +q^u\left[v_S\cosh s-v_B\cosh(s-|x_B|)\right]^2\,{\rm csch}^2\,x_B.
\fe

\section{Semiclassical block and worldline action in Euclidean signature}\label{sec:semiwline}

In this appendix, we consider the Euclidean semiclassical Virasoro blocks with spinning intermediate operators (${\mathfrak s}=h-\bar h\neq 0$), and match them with the on-shell worldline action of spinning particles. We consider scalar external operators $h_1=\bar h_1=h_2=\bar h_2=h_v$ and $h_3=\bar h_3=h_4=\bar h_4=h_w$, and focus on the following two limits: 
\begin{enumerate}
\item Heavy-light limit: ${h_v\over c},{h\over c},{\bar h\over c}\to 0$ with ${h_w\over c}$ fixed. 
\item Light limit: ${h_v\over c},{h_w\over c},{h\over c},{\bar h\over c}\to 0$.
\end{enumerate}

\subsection{Heavy-light limit}

The semiclassical heavy-light limit of the Virasoro block \eqref{eqn:heavyLightBlock} corresponds holographically to a worldline action on the background of a conical defect \cite{Hijano:2015rla}. The pair of heavy operators $W$'s create a conical defect geometry, and the bulk dual of the pair of light operators $V$'s is a particle propagating in the background of the conical defect, whose worldline action is equal to the semiclassical heavy-light block \eqref{eqn:heavyLightBlock}. The trajectory of the particles consists of three segments. The first two segments are geodesics from the two boundary points of the operators $V$'s to a point in the bulk, denoted by ${\mathscr B}$. The third segment is a geodesic from the bulk point ${\mathscr B}$ to a point on the conical singularity, denoted by ${\mathscr C}$. This relation between the semiclassical heavy-light block and worldline action has been explicitly verified for the case $h=\bar h$ in \cite{Hijano:2015rla}, where the intermediate particle is a scalar and the worldline action of it is simply given by its geodesic distance. We  extend this check to the case $h\neq\bar h$.
 
Consider the global patch,
\ie
ds^2={1\over \cos^2\rho}d\rho^2+{\A^2\over\cos^2\rho }d\tau^2+\A^2\tan^2\rho d\phi^2,
\fe
where the coordinates are in the range $\rho\in[0,{\pi\over 2}),\,\tau\in(-\infty,\infty),\,\phi\in[0,2\pi)$. The conical singularity is at $\rho=0$ with deficit angle $2\pi(1-\A)$. We introduce the complex coordinate 
\ie
w=\phi+i\tau.
\fe
One of the $V$'s is inserted on the boundary at $w=0$, and the other $V$ is inserted at general $w$.  The CFT cross ratio is related to the coordinate $w$ by
\ie
1-z=e^{iw}.
\fe
The semiclassical heavy-light block \eqref{eqn:heavyLightBlock} in the variable $w$ is
\ie
z^{2h_w}{\cal F}(h,z)
&=\left({\A\sin{w\over 2}\over \sin{\A w\over 2}}\right)^{2h_v}\left(-{4i\over \A}\tan\tfrac{\A w}{4}\right)^h.
\fe
The holomorphic and anti-holomorphic blocks combine to give
\ie
z^{2h_w}\bar z^{2h_w}{\cal F}(h,z){\cal F}(\bar h,\bar z)
&=\left({\A^2\sin{ w\over 2}\sin{\bar w\over 2}\over \sin{\A w\over 2}\sin{\A \bar w\over 2}}\right)^{2h_v}\left({4\over \A}\tan\tfrac{\A w}{4}\right)^h\left({4\over \A}\tan\tfrac{\A \bar w}{4}\right)^{\bar h}e^{-i{\pi\over 2}( h-\bar h)}.
\fe
Let us specialize to the case that the operators $V$'s are on a constant time slice $\tau=0$. As a consequence, the three segments of geodesics are all on the slice $\tau=0$, and the geodesic $\overline{{\mathscr B} {\mathscr C}}$ is along the $\rho$ direction. The heavy-light block with $w=\bar w=\phi$ is given by
\ie\label{eqn:EuclideanFFw=bw}
z^{2h_w}\bar z^{2h_w}{\cal F}(h,z){\cal F}(\bar h,\bar z)\Big|_{w=\bar w}
&=\left({\A\sin{ \phi\over 2}\over \sin{\A \phi\over 2}}\right)^{4h_v}\left({4\over \A}\tan\tfrac{\A \phi}{4}\right)^{h+\bar h}e^{-i{\pi\over 2}( h-\bar h)}.
\fe
The log of the modulus of $z^{2h_w}\bar z^{2h_w}{\cal F}(h,z){\cal F}(\bar h,\bar z)|_{w=\bar w}$ is equal to the sum of the geodesic distances of the three segments of worldlines \cite{Hijano:2015rla}. The phase of $z^{2h_w}\bar z^{2h_w}{\cal F}(h,z){\cal F}(\bar h,\bar z)|_{w=\bar w}$ corresponds to the spinning particle action $S_{spin}$ in \eqref{eqn:SspinEta} with $\eta=i\theta$ for Euclidean signature. According to our prescription, the normal vector $\tilde n$ at an interaction vertex is in the two-plane spanned by the velocity vectors of the vertex.  Hence, the normal vector $\tilde n$ at the bulk point ${\mathscr B}$ (bulk point ${\mathscr C}$) is in the $\phi$-direction ($\tau$-direction). Parallel transporting from ${\mathscr B}$ to ${\mathscr C}$ along the $\rho$-direction would not change $\tilde n$. Hence, the action $S_{spin}$ is simply given by
\ie
S_{spin}=i(h-\bar h){\pi\over 2},
\fe
and $e^{-S_{spin}}$ precisely gives the phase of $z^{2h_w}\bar z^{2h_w}{\cal F}(h,z){\cal F}(\bar h,\bar z)|_{w=\bar w}$ in \eqref{eqn:EuclideanFFw=bw}.

\subsection{Light limit}

Let us parametrize the following configuration of the four-point function 
\ie\label{eqn:FBPTs}
z_1=\rho,~~~z_2=-\rho,~~~z_3=1,~~~z_4=-1.
\fe
The cross ratio is
\ie
z={4\rho^2\over (1+\rho)^2}.
\fe
The semiclassical block in the light limit takes a simple form as
\ie\label{eqn:heavyLightBlockSmallHw}
{\cal F}(h,z)
&=z^{-2h_w}(4\rho)^h,~~~\rho={z\over (1+\sqrt{1-z})^2}.
\fe
Plugging \eqref{eqn:heavyLightBlockSmallHw} into \eqref{eqn:virasoroBlockDecomp}, we find
\ie\label{eqn:GSmallHw}
G(z,\bar z)=\sum_{a} C_{WW{\cal O}_a} \, C_{VV{\cal O}_a}\, (4\rho)^{h}\,(4\bar \rho)^{\bar h}.
\fe

The factor $(4\rho)^{h}(4\bar \rho)^{\bar h}$ is equal to an on-shell worldline action in Euclidean AdS$_3$. We will work in the Poincar\'e patch,
\ie
ds^2={dy^2+dzd\bar z\over y^2},
\fe
where the boundary is at $y=0$. Let us introduce two bulk points $\vec{\bf x}_a = (z=0, y=y_a )$ and $\vec{\bf x}_b = (0, y_b )$. Consider two geodesics from the boundary points $z_1$ and $z_2$ to the first bulk point $\vec{\bf x}_a$, and another two geodesics from the boundary points $z_3$ and $z_4$ to the second bulk point $\vec{\bf x}_b$. We introduce one more geodesic from the bulk point $\vec{\bf x}_a$ to the bulk point $\vec{\bf x}_b$. The worldline action is given by
\ie
S&=S_{ wl}+S_{spin},
\\
S_{ wl}&=2h_v\left[ L( z_1,\vec{\bf x}_a)+ L( z_2,\vec{\bf x}_a)+ L( z_3,\vec{\bf x}_b)+ L( z_4,\vec{\bf x}_b)\right]+(h+\bar h)L(\vec{\bf x}_a,\vec{\bf x}_b),
\fe
where $S_{spin}$ is given by \eqref{eqn:SspinEta}, and for Euclidean signature we have $\eta=i\theta$,
\ie
S_{spin}&=i(h-\bar h)(\theta_f-\theta_i).
\fe
Using \eqref{eqn:geodesicDistance} and \eqref{eqn:bdryBulkGeodesic}, the $S_{ wl}$ is given explicitly by
\ie\label{eqn:Swl}
S_{ wl}=4h_v\left[\log\left({y^2_a+|\rho|^2\over 2|\rho| y_a}\right)+\log\left({y^2_b+1\over 2 z_b}\right)\right]+(h+\bar h)\cosh^{-1}\left[1+{(y_a-y_b)^2\over 2y_a y_b}\right],
\fe
which is minimized at
\ie
z_a=|\rho| \sqrt{2h_v+(h+\bar h)\over 2h_v-(h+\bar h)},~~~z_b=\sqrt{2h_v-(h+\bar h)\over 2h_v+(h+\bar h)}.
\fe
Plugging this back into \eqref{eqn:Swl}, we have
\ie
e^{-S_{wl}}=|\rho|^{h+\bar h}\left(4h_v^2-(h+\bar h)^2\over 4h_v^2\right)^{4h_v}\left(2h_v+(h+\bar h)\over 2h_v-(h+\bar h)\right)^{2h}.
\fe
To determine the $S_{spin}$, let us denote the difference between the velocity vectors of the geodesics $\overline{z_1 \vec{\bf x}_a}$ and $\overline{z_2 \vec{\bf x}_a}$ at $\vec{\bf x}_a$ by $\Delta v_a$, and the difference between the velocity vectors of the geodesics $\overline{z_3 \vec{\bf x}_b}$ and $\overline{z_4 \vec{\bf x}_b}$ at $\vec{\bf x}_b$ by $\Delta v_b$. The normal vector $\tilde n$ at $\vec{\bf x}_a$ (or $\vec{\bf x}_b$) is proportional to $\Delta v_a$ (or $\Delta v_b$). Since $\Delta v_a$ and $\Delta v_b$ has no component in the $y$-direction, the parallel transport along the $y$-direction would not change $\Delta v_a$ and $\Delta v_b$. The $\theta_f-\theta_i$ is the angle between $\Delta v_a$ and $\Delta v_b$, and we have
\ie
S_{spin}=(h-\bar h)\log\left({\rho\over |\rho|}\right).
\fe
Putting everything together, we find
\ie
e^{-S}\propto (4\rho)^{h}\,(4\bar \rho)^{\bar h}.
\fe


\providecommand{\href}[2]{#2}\begingroup\raggedright\endgroup

\end{document}